\documentclass[onecolumn,preprintnumbers,amsmath,amssymb,aps]{revtex4}

 \normalsize

\usepackage[dvips]{color}
\usepackage{psfrag}
\usepackage{graphicx,graphics}
\usepackage{dcolumn}
\usepackage{bm}
\usepackage{amssymb}
\usepackage{slashed}
\usepackage{amsmath}
\usepackage{float}
\usepackage{pdfpages}
\graphicspath{{./Figures/}}
\newtheorem{theorem}{Theorem}

\def\bea{\begin{eqnarray}}
\def\eea{\end{eqnarray}}
\def\beas{\begin{eqnarray*}}
\def\eeas{\end{eqnarray*}}
\def\be{\begin{equation}}
\def\ee{\end{equation}}
\def\bes{\begin{equation*}}
\def\ees{\end{equation*}}
\def\lsim{\raise0.3ex\hbox{$\;<$\kern-0.75em\raise-1.1ex\hbox{$\sim\;$}}}
\def\gsim{\raise0.3ex\hbox{$\;>$\kern-0.75em\raise-1.1ex\hbox{$\sim\;$}}}

\def\nn{\nonumber}
\def\bt{\begin{table}}
\def\et{\end{table}}

\newcommand{\lm}{\lambda}
\newcommand{\tw}{\theta_w}
\newcommand{\al}{\alpha}

\begin{document}
\title{Phenomenological Aspects of a TeV Scale Alternative left-right Model}
\author{M. Ashry$^{1,2}$ and S. Khalil$^{2,3}$}
\affiliation{
$^1$Department of Mathematics,~Faculty of Science,~Cairo University,~12613,~Giza,~Egypt.\\
$^2$Center for Fundamental Physics,~Zewail City of Science and Technology,~Sheikh Zayed,~12588,~Giza,~Egypt.\\
$^3$Department of Mathematics,~Faculty of Science,~Ain Shams University,~11566,~Cairo,~Egypt.}
\begin{abstract}
We revisit the alternative left-right symmetric model, motivated by the superstring-inspired $E_6$ model. We systematically analyze the constraints imposed by theoretical and experimental bounds on the parameter space of this class of models. We perform a comprehensive analysis of the Higgs sector and show that three neutral $CP$-even and two $CP$-odd Higgs bosons in addition to two charged Higgs bosons can be light, of ${\cal O}(100)$ GeV.  We emphasize that the predictions of this model for the signal strengths of Higgs decays are consistent with the standard model expectations. We also explore discovery signatures of the exotic down-type quark, which is one of the salient predictions of this model.
\end{abstract}
\maketitle
\section{Introduction}
The discovery of neutrino masses and oscillations confirmed the
fact that although the standard model (SM) is extremely accurate,
it is still incomplete. The left-right Model (LRM) is the most
natural extension of the SM that accounts for the measured
neutrino masses and provides an elegant understanding for the
origin of the parity violation in low-energy weak interactions
\cite{Mohapatra:1974gc,Senjanovic:1975rk,Mohapatra:1977mj,Deshpande:1990ip,Aulakh:1998nn,Maiezza:2010ic,Borah:2010zq,Nemevsek:2012iq}.
The LRM is based on the gauge group $SU(3)_C \times
SU(2)_L \times SU(2)_R \times U(1)_{(B-L)/2}\times P$, where $P$ is
the discrete parity symmetry. In the LRM, standard model fermions
are assigned in the following left- or right-handed doublets:
\be
Q_L\equiv\left(\begin{array}{c}u_L\\d_L\end{array}\right),~\psi_L\equiv\left(\begin{array}{c}\nu_L\\e_L\end{array}\right)
~~~\& ~~~
Q_R\equiv\left(\begin{array}{c}u_R\\d_R\end{array}\right),~\psi_R\equiv\left(\begin{array}{c}\nu_R\\e_R\end{array}\right).
\ee
The parity symmetry: $Q_L, \psi_L \leftrightarrow Q_R, \psi_R$ implies that the gauge couplings of left- and right-handed $SU(2)$ are
equal, {\it i.e.}, $g_L = g_R \equiv g$.

The Higgs sector of the LRM consists of $(i)$ bidoublet
$\Phi(1,2,2^*,0)$, which is required to construct the SM Yukawa
couplings of quarks and leptons. $(ii)$ two scalar triplets
$\Delta_L(1,3,0,2)$ and $\Delta_R(1,0,3,2)$ that break
$U(1)_{(B-L)/2}$ and generate neutrino Majorana masses. At high-energy
scale, well above the electroweak breaking scale, the $SU(2)_R
\times U(1)_{(B-L)/2} \times P$ symmetry is broken down to $U(1)_Y$ by
the vacuum expectation value (vev) of the neutral component of
$\Delta_R$, and hence the right-handed Majorana neutrino mass is
generated. In this type of models, the hypercharge $Y$ is defined
as $ Y = T_{3R} + (B - L)/2$, where $T_{3R}$ is the third
component of the right-handed isospin. At lower energy scales,
$\Phi$ and $\Delta_L$ acquire vevs that break $SU(2)_L \times
U(1)_Y$ down to $U(1)_{em}$. It is worth mentioning that in
the conventional LRM, one gets the following estimate for the
associated vevs: $\langle \Delta_L \rangle = v_L \lsim {\cal
O}(1)$ GeV, $\langle \Delta_R \rangle = v_R \gsim {\cal
O}(10^{11})$ GeV, and $\langle \Phi \rangle = {\rm diag}\{\kappa,
\kappa'\}$ with $\kappa' \ll \kappa$ and $\kappa \sim {\cal
O}(100)$ GeV \cite{Mohapatra:1974gc,Mohapatra:1977mj,Deshpande:1990ip}.

It turns out that the Higgs sector of the LRM, in particular the Higgs triplets, may
induce tree-level flavor-violating processes that contradict the
current experimental limits. Therefore, it is usually assumed that
$SU(2)_R \times U(1)_{(B-L)/2}$ is broken at a very high-energy scale.
In this case, it is not possible to detect any residual effect for
$SU(2)_R$ gauge symmetry at the TeV scale in the Large Hadron Collider
(LHC). This has motivated Ernest Ma, in his pioneering work in
1987 \cite{Ma:1986we}, to study variants of the conventional LRM.
He has shown that the superstring-inspired $E_6$ model may lead to
two types of left-right models. The first one is the canonical LRM,
while the second one is what is known as the alternative left-right Model (ALRM) \cite{Babu:1987kp,Ma:2010us}, where the fermion assignments are
different from those in the conventional LRM in the following: $(i)$ an extra quark, $d'_R$, instead of
$d_R$, is combined with $u_R$ and form $SU(2)_R$ doublet, and$(ii)$ an extra lepton, $n_R$, instead of $\nu_R$, is combined with $e_R$ and
forms $SU(2)_R$ doublet. Therefore, the right-handed neutrino
$\nu_R$ is a true singlet and is no longer a part of the
right-handed doublet.

It is remarkable that $E_6$ is a complex Lie group of rank 6. It includes the $SO(10)$ group,
so it is a good candidate for grand unification. Some string theories (Heterotic string) predict
that the low-energy effective model is symmetric under $E_6$. Depending on the string model,
$E_6$ may be broken to $SO(10)$ and then to the conventional left-right model or it may have
another branch of symmetry breaking that leads to the alternative left-right model that we consider.
The particle content of the ALRM, derived
from $E_6$ model, contains more particles than those in the
conventional LRM obtained from $SO(10)$. This can be simply
understood from the fact that the fundamental representation $27$
of $E_6$ is equivalent to the fundamental representation $16$ of
$SO(10)$ plus its $10$ and singlet representations. In the
conventional LRM, all non-SM particles are decoupled and can be
quite heavy. However, in the ALRM, they are involved with the SM
fermions and will have low-energy consequences. Furthermore,
another important difference between the ALRM and the conventional
LRM is the fact that tree-level flavor-changing neutral currents
are naturally absent so that the $SU(2)_R$ breaking scale can be
of order TeV, allowing to several interesting signatures at the
LHC. As the ALRM is a low-energy effective model of the supersymmetric $E_6$ model,
the gauge couplings are not unified within the ALRM. They are unified in the
underling $E_6$ model, similar to the unification of SM gauge couplings in supersymmetric $SU(5)$.

In this paper, we aim at providing a comprehensive analysis for the
phenomenological implications of the ALRM, with emphasis on the
possible signatures of this model at the LHC. There are couple of
recent papers \cite{Ma:2010us,Khalil:2010yt} that discuss specific phenomenological
aspects of the ALRM, namely, the dark matter search and $Z'$ and
$W'$ signals at the LHC. Our goal here is twofold. The first one isto
analyze the Higgs sector of the ALRM and check if the recent
results reported by ATLAS and CMS experiments on Higgs production
and decays can be accommodated. The Second is to explore the discovery
signature of the exotic down-type quarks associated with this type
of models at the LHC.

The latest results of ATLAS and CMS collaborations \cite{Aad:2014eha,Khachatryan:2014ira},
confirmed the Higgs discovery with mass around 125 GeV, through
Higgs decay channels: $H \to \gamma \gamma$, $H \to ZZ^{(*)} \to 4
l$, and $H \to WW^{(*)} \to l \nu l \nu$ at integrated
luminosities of $5.1$ fb$^{-1}$ taken at energy $\sqrt{s} =7$ TeV
and $19.6$ fb$^{-1}$ taken at $\sqrt{s}=8$ TeV.  These results still indicate
possible discrepancies between their results for signal
strengths in these channels \cite{Moriond1-CMS:2013,CMSNOTE1:2013,Moriond2-ATLAS:2013,ATLASNOTE2:2013}. We show that our ALRM has a
rich Higgs sector, and consists of one bidoublet and two left-handed
and right-handed doublets. Therefore, one obtains four neutral
$CP$-even and two $CP$-odd Higgs bosons, in addition to two charged Higgs bosons.
It turns out that most of these Higgs bosons can be light, of the order
the electroweak scale, and can be accessible at the LHC. We also find that
the contributions of the charged Higgs bosons to the decay rate of $H \to \gamma \gamma$
are not significant. Furthermore, we show that, due to the mixing
among the neutral $CP$-even Higgs bosons, the couplings of the
SM-like Higgs, which is the lightest one, with the top quark and
$W$-gauge boson are slightly modified respect to the SM ones.
Therefore, the ALRM predictions for signal strengths of Higgs
decays, in particular, $H \to \gamma \gamma$ and $H\to W^+ W^-$,
are consistent with the SM expectation.

Another salient feature of the ALRM is the presence of an extra
down-type quark, $d'$. We analyze the striking signature of this
exotic quark at the LHC. We show that the most promising
$d'$-production channel is $g g \to \bar{d}' d'$, due to the
direct coupling of $d'$ to gluons with a strong coupling constant
and color factor. Then, $d'$ decays to a jet and lepton plus missing
energy. We find that the cross section of this process is of
${\cal O}(1)$ fb, which can be probed at the LHC with 14 TeV
center-of-mass energy.

The paper is organized as follows. In Sec. 2, we briefly review
the TeV scale ALRM. Section 3 is devoted to the Higgs sector, in particular, for
studying the mixing matrix of the Higgs bosons and investigating the
existence of two light charged Higgs bosons. In Sec. 4 we focus
on the Higgs decay into a diphoton in the ALRM.
The discovery signatures of extra quark $d'$ at the LHC is discussed in Sec. 5.
Finally, we give our conclusions in Sec. 6.

\section{Alternative left-right Symmetric Model}
We consider an ALRM based on $SU(3)_C \times SU(2)_L \times
SU(2)_R \times U(1)_{(B-L)/2} \times S$, where $S$ is a discrete
symmetry imposed to distinguish between scalars and their
dual scalars. The fermion content of this model, with their
charge assignments, is presented in Table~\ref{pcont} \cite{PhysRevD.85.091701}. As
can be seen from this table, extra quarks and leptons are
predicted as in all $E_6$-based left-right models.
\bt[t]
\centering {\begin{tabular}{|ccc|}\hline
\textbf{Fields}&$SU(3)_c\times SU(2)_L\times SU(2)_R\times U(1)_{(B-L)/2}$&$S$\\\hline
\textbf{Fermions}&&\\
$Q_L\equiv\left(\begin{array}{c}u\\d\end{array}\right)_L$&$(3,2,1,+\frac{1}{6})$&0\\
$Q_R\equiv\left(\begin{array}{c}u\\d'\end{array}\right)_R$&$(3,1,2,+\frac{1}{6})$&$-\frac{1}{2}$\\
$d'_L$&$(3,1,1,-\frac{1}{3})$&$-1$\\
$d_R$&$(3,1,1,-\frac{1}{3})$&0\\
$\psi_L\equiv\left(\begin{array}{c}\nu\\e\end{array}\right)_L$&$(1,2,1,-\frac{1}{2})$&0\\
$\psi_R\equiv\left(\begin{array}{c}n\\e\end{array}\right)_R$&$(1,1,2,-\frac{1}{2})$&$+\frac{1}{2}$\\
$n_L$&$(1,1,1,0)$&$+1$\\
$\nu_R$&$(1,1,1,0)$&$0$\\\hline
\textbf{Higgs}&&\\
$\Phi\equiv\left(\begin{array}{cc}\phi_1^0&\phi_1^+\\ \phi_2^-&\phi_2^0\end{array}\right)$&$(1,2,2^*,0)$&$-\frac{1}{2}$\\
$\chi_L\equiv\left(\begin{array}{c}\chi_L^+\\ \chi_L^0\end{array}\right)$&$(1,2,1,+\frac{1}{2})$&0\\
$\chi_R\equiv\left(\begin{array}{c}
\chi_R^+\\ \chi_R^0\end{array}\right)$&$(1,1,2,+\frac{1}{2})$&$+\frac{1}{2}$\\\hline
\end{tabular}}
\caption{Particle content and their quantum numbers in the ALRM.}
\label{pcont}
\et

The Higgs sector of our ALRM consists of an $SU(2)_R$ scalar doublet
$\chi_R$ to break $SU(2)_R \times U(1)_{(B-L)/2}$ in addition to
$SU(2)_L$ scalar doublet $\chi_L$ and scalar bidoublet $\Phi$
that break $SU(2)_L \times U(1)_Y$. The detailed quantum numbers
of these Higgs bosons are presented in
Table~\ref{pcont} \cite{PhysRevD.85.091701}.

In this case, the most general left-right symmetric Yukawa Lagrangian is given by%
\bea
\mathcal{L}_\text{Y}&=&
\overline{Q}_LY^q\widetilde{\Phi}Q_R+\overline{Q}_L Y^q_L\chi_L d_R+\overline{Q}_R Y^q_R\chi_R d'_L+\overline{\psi}_L Y^\ell\Phi\psi_R\nn
\\&+&\overline{\psi}_L Y^\ell_L\widetilde{\chi}_L\nu_R+\overline{\psi}_R Y^\ell_R\widetilde{\chi}_R n_L
+\overline{\nu}^c_R M_R\nu_R + \text{h.c.}\ ,
\label{Yukawa}
\eea
where $\widetilde{\Phi}$ is the dual of the scalar bidoublet
$\Phi$, defined as $\widetilde{\Phi}=\tau_2\Phi^*\tau_2$, and
$\widetilde{\chi}_{L,R}$ are the dual of the scalar doublets $\chi_{L,R}$,
defined as $\widetilde{\chi}_{L,R}=i\tau_2\chi^*_{L,R}$. Note
that the Yukawa terms like $\overline{\psi}_L \tilde{\Phi} \psi_R$ and
$\overline{Q}_L \Phi Q_R$ are forbidden by the discrete $S$ symmetry
only. A detailed discussion on the Higgs potential and the
associated vevs will be given in the next section. Here, we assume
a nonvanishing vev of $\chi_R$, $\langle \chi_R \rangle =
v_R/\sqrt{2}$ of order TeV with vevs of $\chi_L$ and $\Phi$, given
by $\langle \chi_L \rangle = v_L/\sqrt{2}$ and $\langle \Phi
\rangle = {\rm diag}\{0,k/\sqrt{2}\}$. The breaking of $SU(2)_R
\times U(1)_{(B-L)/2}$ down to $U(1)_Y$ leaves the discrete symmetry:  $L= S +T_{3R}$ unbroken, if the vev
of $\phi^0_1$ (which has $L=-1$) is zero, while $\phi^0_2$, (with $L=0$) could have a nonvanishing vev. In this case, one can
easily show that the quarks $u$, $d$ and $d'$ and the
charged leptons $\ell$, in addition to the singlet fermion $n$, which is called a scotino, acquire the following masses
\bea
m_u&=&\frac{1}{\sqrt{2}}Y^q v\sin\beta,~~
m_d=\frac{1}{\sqrt{2}}Y^q_L v\cos\beta,~~
m_{d'}=\frac{1}{\sqrt{2}}Y^q_R v_R,\\
m_\ell&=&\frac{1}{\sqrt{2}}Y^\ell v\sin\beta,~~
m_{n}=\frac{1}{\sqrt{2}}Y^\ell_R v_R,
\eea
where $\tan\beta=k/v_L$ and $\sqrt{v_L^2 + k^2} = v \equiv 246 ~ {\rm GeV}$. Moreover, the neutrino mass matrix is given by
\be
M_\nu=\left(\begin{array}{c|cc}
    &\nu^c_L&\nu_R\\\hline
    \overline{\nu}_L&0&m_{\nu D}\\
    \overline{\nu}^c_R&m_{\nu D}^T&M_R
\end{array}\right),
\ee
where  $m_{\nu D}=Y^\ell_Lv_L/\sqrt{2}$ . The mass $M_R$ is not
related to the $SU(2)_R$ symmetry-breaking scale, so it can be quite large. This matrix can be diagonalized by a unitary
matrix, $V_\nu M_\nu V_\nu^T\simeq \text{diag}\left(m_{\nu_l},m_{\nu_h}\right)$,
where $m_{\nu_l}$ and $m_{\nu_h}$ are the well known seesaw mass eigenvalues of the light and heavy neutrinos, respectively:
\be
m_{\nu_l} \simeq m_{\nu D}M_R^{-1}m_{\nu D}^T,~~~~~ m_{\nu_h} \simeq M_R.
\ee

Now, we turn to the gauge sector of the ALRM; the covariant derivatives of the Higgs bosons are given by
\bea
D_\mu\Phi&=&\partial_\mu\Phi-i\frac{g}{2}\left(\tau^aW^a_{L_\mu}~\Phi-\Phi~\tau^a W^a_{R_\mu}\right),\\
D_\mu\chi_{L,R}&=&\partial_\mu\chi_{L,R}-i\frac{g}{2}\tau^aW^a_{{L,R}_\mu}\chi_{L,R}-i\frac{g_{BL}}{2}B_\mu\chi_{L,R}~,
\eea
where $g_{BL}$ is the gauge coupling of the $U(1)_{(B-L)/2}$ group. After the spontaneous breaking of left-right symmetry down to
electroweak and then down to electromagnetism, the associated gauge
bosons acquire masses, through the nonvanishing vevs of $\chi_R$,
$\Phi$, and $\chi_L$.
Because of the vanishing vev of $\phi_1^0 \in\Phi$, the mixing between $W^\pm_L$ and $W^\mp_R$ is identically zero. Thus, the physical eigenstates are given by SM gauge bosons $W^\pm=W_L^\pm$ and $W'^\pm=W_R^\pm$ with masses
\bea%
M_W^2&=&\frac{1}{4}g^2\left(k^2+v_L^2\right)=\frac{1}{4}g^2 v^2,\\
M_{W'}^2&=&\frac{1}{4}g^2\left( k^2+v_R^2\right).
\eea
The experimental searches for $W'$ at the LHC through their decays to
electron/muon and neutrino lead to $M_{W'} \gsim 2.5 $ TeV
\cite{Chatrchyan:2012meb,Aad:2011yg}. The interactions of our
$W'$ with the SM fermions are given by%
\bea
\mathcal{L}_{\text{gauge}}^{W'} &=&
-\frac{i g}{\sqrt{2}}\overline{u}~\gamma^\mu {W'}_\mu^+ V'_{\text{CKM}}P_R~d'-\frac{i g}{\sqrt{2}}\overline{d}'~\gamma^\mu {W'}_\mu^- {V'}_{\text{CKM}}^\dag P_R~u\nn\\
&&-\frac{i g}{\sqrt{2}}\overline{n}~\gamma^\mu {W'}_\mu^+ U'_{\text{MNS}} P_R~e -\frac{i g}{\sqrt{2}}\overline{e}~\gamma^\mu {W'}_\mu^-{U'}_{\text{MNS}}^\dag P_R~n. %
\eea
Thus, $W'$ can decay into an electron and singlet fermion
(scotino) $n$, which appears at the LHC as missing energy.
Therefore, the above-mentioned lower bound on $M_{W'}$ is
applicable in our ALRM. This implies that $v_R \gsim {\cal O}(1)$
TeV. The situation of the neutral gauge bosons: $W_L^3, W_R^3$ and $B$
is more involved. One can show that their mass matrix is given by
\be
\left(\begin{array}{c|ccc}&W_L^3&W_R^3&B\\\hline
W_L^3&\frac{1}{4}g^2\left( k^2+v_L^2\right)&-\frac{1}{4}g^2 k^2&-\frac{1}{4}gg_{BL} v_L^2\\
W_R^3&-\frac{1}{4}g^2 k^2&\frac{1}{4}g^2\left( k^2+v_R^2\right)&-\frac{1}{4}g g_{BL}  v_R^2\\
B&-\frac{1}{4}g g_{BL} v_L^2&-\frac{1}{4}g g_{BL} v_R^2&\frac{1}{4}{g_{BL}}^2\left(v_L^2+v_R^2\right)\end{array}
\right).
\ee
One can define $s_{w}\equiv\sin\theta_w=e/g$, and with $c_w\equiv\cos\theta_{w}$, then
$g_{BL} =e/\sqrt{c_{w}^2-s_{w}^2}$. It is more convenient to work in
the basis $(A,Z_L,Z_R)$, where
\be\left(\begin{array}{c}A\\Z_L\\Z_R\end{array}\right)=
\left(\begin{array}{ccc}
s_{w}&s_{w}&\sqrt{c_{w}^2-s_{w}^2}\\
c_{w}&-s_{w}^2/c_{w}&-s_{w}\sqrt{c_{w}^2-s_{w}^2}/c_{w}\\
0&\sqrt{c_{w}^2-s_{w}^2}/c_{w}&-s_{w}/c_{w}
\end{array} \right)
\left(\begin{array}{c}W_L^3\\W_R^3\\B\end{array}\right). \ee

In this case, one can show that the mass eigenvalue of the gauge boson $A$ is identically zero.
Therefore this gauge boson is the photon that should remain massless after symmetry breaking. The exact eigenstates $Z,Z'$ are obtained
as
\be
\left(\begin{array}{c}Z\\Z'\end{array}\right)=\left(\begin{array}{cc}
\cos\vartheta& \sin\vartheta\\
-\sin\vartheta&\cos\vartheta\end{array} \right)
\left(\begin{array}{c}Z_L\\Z_R\end{array}\right).
\ee
The mixing angle $\vartheta$ is defined as
\be
\tan2\vartheta=\frac{2M_{LR}^2}{M_{LL}^2-M_{RR}^2},\label{Z-Z' Mixing}
\ee
where
\bea
M_{LL}^2&=&\frac{g^2 v^2}{4\cos^2\theta_w},\\
M_{LR}^2&=&\frac{g^2 (v^2\sin^2\theta_w-k^2 \cos^2\theta_w)}{4\cos^2\theta_w\sqrt{\cos2\theta_w}},\\
M_{RR}^2&=&\frac{g^2 (2v^2\sin^4\theta_w+2(k^2+v_R^2) \cos^4\theta_w-k^2\sin^22\theta_w)}{8\cos^2\theta_w \cos2\theta_w}.
\eea
The eigenvalues $M_Z^2$ and $M_{Z'}^2$ are given by
\be
M_{Z,Z'}^2=\frac{1}{2}\left(M_{LL}^2+M_{RR}^2\mp\left(M_{RR}^2-M_{LL}^2\right)\sqrt{1+\tan^22\vartheta}\right).
\ee
It is clear that if $v_R \gg v$, {\it i.e.}, $\vartheta \to 0$, then
$Z\simeq Z_L$ and $Z' \simeq Z_R$. The LHC search for the $Z'$ gauge
boson is rather model dependent. However, one may consider $M_{Z'}
\gsim 2$ TeV as a conservative lower bound \cite{Collaboration:2011dca,Chatrchyan:2011wq}. In addition, the
mixing between $Z$ and $Z'$ should be less than ${\cal O}(10^{-3})$.
\section{Higgs Sector in the ALRM}
\subsection{Symmetry breaking}
%
The Higgs sector of our ALRM consists of bidoublet $\Phi$ with
left and right doublets $\chi_L$ and $\chi_R$. The charge
assignments of these Higgs bosons are shown in Table
\ref{pcont}. As mentioned in the previous section, the gauge
symmetries $SU(2)_R \times U(1)_{(B-L)/2}$ are spontaneously broken to
$U(1)_Y$ through the vev of $\chi_R$, and then $SU(2)_L \times U(1)_Y$
symmetries are broken by vevs of $\Phi$ and $\chi_L$. The most
general Higgs potential that is invariant under these symmetries is given
by \cite{Borah:2010zq}
\bea
V(\Phi,\chi_{L,R})&=&
-\mu_1^2Tr[\Phi^\dag\Phi]+\lm_1(Tr[\Phi^\dag\Phi])^2+\lm_2 Tr[\Phi^\dag\widetilde{\Phi}] ~ Tr[\widetilde{\Phi}^\dag\Phi] \nn \\&~&-\mu_2^2(\chi_L^\dag\chi_L+\chi_R^\dag\chi_R)+\lm_3[(\chi_L^\dag\chi_L)^2+(\chi_R^\dag\chi_R)^2]+2\lm_4(\chi_L^\dag\chi_L)(\chi_R^\dag\chi_R)\nn
\\&~&+2\al_1Tr(\Phi^\dag\Phi)(\chi_L^\dag\chi_L+\chi_R^\dag\chi_R)+2\al_2(\chi_L^\dag\Phi\Phi^\dag\chi_L+\chi_R^\dag\Phi^\dag\Phi\chi_R)\nn
\\&~&+2\al_3(\chi_L^\dag\widetilde{\Phi}\widetilde{\Phi}^\dag\chi_L+\chi_R^\dag\widetilde{\Phi}^\dag\widetilde{\Phi}\chi_R)
+\mu_3(\chi_L^\dag\Phi\chi_R+\chi_R^\dag\Phi^\dag\chi_L).~~~~~~
\label{Scalar Potential}
\eea%
In the Appendix, we provide a detailed study for the conditions that keep the potential (\ref{Scalar Potential}) bounded from below. It is remarkable that the copositivity conditions \cite{Ping1993109,Kannike:2012pe} for this Higgs potential significantly depend on the signs of the following parameters $\al_{12}=\al_1+\al_2,\ \al_{13}=\al_1+\al_3,$ and $\lm_{12}=\lm_1+2\lm_2$. Here, we present the case with minimal constraints imposed on the potential parameters:
\be\label{stcond} \lm_1\geq0,~\lm_2\leq0,~\lm_3\geq0,~\al_2-\al_3\geq0,~\al_{12}\geq0,\ \al_{13}\geq0,\ \lm_{12}\geq0. \ee
Also for perturbativity the absolute value of any dimensionless potential parameter is assumed to be less than $\sqrt{4\pi}$.
In addition, from the minimization conditions, one finds that the nonvanishing vevs are given by
\bea
v_L v_R&=&\frac{-\mu_3 k}{\sqrt{2}\left(\lm_4-\lm_3\right)},\label{mca}\\
v_L^2+v_R^2&=&\frac{\mu_2^2-\al_{12} k^2}{\lm_3},\label{mcb}\\
k^2&=&\frac{2\left(\lm_3\mu_1^2-\al_{12}\mu_2^2\right)\left(\lm_4-\lm_3\right)+
\lm_3\mu_3^2}{2\left(\lm_1\lm_3-\al_{12}^2\right)\left(\lm_4-\lm_3\right)}.
\label{mcc}
\eea
We use these equations to determine three parameters ($\mu_1$,
$\mu_2$, and $\lm_4$) out of the ten free parameters in the Higgs
potential (\ref{Scalar Potential}) in terms of the vevs: $k=v\sin\beta,\ v_L = v\cos\beta$, and $v_R\sim{\cal O}(1)$ TeV. Note that since the vevs $k,v_L$ are of the same order and the couplings $\lm_{3,4} \lsim {\cal O}(1)$, the values of $\mu_3$ can be  smaller than $v_R$.

\subsection{Higgs masses and mixing}
We begin by $16$ degrees of freedom; $8$ of $\Phi$ and $8$
of $\chi_L$ and $\chi_R$. After symmetry breaking, two neutral
components of these $16$ degrees of freedom will be eaten by
the neutral gauge bosons $Z$ and $Z'$ to acquire their masses. In
addition, another four charged components will be eaten by the
charged gauge bosons $W^\pm$ and $W'^\pm$ to acquire their masses.
Therefore, ten scalars remain as physical Higgs bosons in this
class of models. As we will explicitly show, four of them give
charged Higgs bosons, two lead to pseudoscalar Higgs
bosons, and the remaining four give $CP$-even neutral Higgs bosons.
\begin{enumerate}
\item \textbf{Charged Higgs bosons}:\\
The mass matrix of the charged Higgs bosons, in the basis
$\left(\begin{array}{cccc}\phi_1^{+} & \chi_L^{+} & \phi_2^+ & \chi_R^{+}\end{array}\right)$, is a block diagonal matrix with the
following two matrices, which, respectively, correspond to the bases
$\left(\begin{array}{cc}\phi_1^{+} & \chi_L^{+}\end{array}\right)$ and $\left(\begin{array}{cc}\phi_2^+ & \chi_R^{+}\end{array}\right)$:
\bea
M_{1L}^+&=&\left(\begin{array}{cc}
 -(\al_2-\al_3)v_L^2-\frac{\mu_3 v_R}{\sqrt{2}}\cot\beta & (\al_2-\al_3)v_L^2\tan\beta-\frac{\mu_3 v_R}{\sqrt{2}}\\
 (\al_2-\al_3)v_L^2\tan\beta-\frac{\mu_3 v_R}{\sqrt{2}} & -(\al_2-\al_3)v_L^2\tan^2\beta-\frac{\mu_3 v_R}{\sqrt{2}}\tan\beta
\end{array}\right),~~~~~\\
&~&\nn\\
M_{2R}^+&=&\left(\begin{array}{cc}
 -(\al_2-\al_3)v_R^2-\frac{\mu_3 v_L}{\sqrt{2}}\cot\zeta & (\al_2-\al_3)v_R^2\tan\zeta-\frac{\mu_3 v_L}{\sqrt{2}}\\
 (\al_2-\al_3)v_R^2\tan\zeta-\frac{\mu_3 v_L}{\sqrt{2}} & -(\al_2-\al_3)v_R^2\tan^2\zeta-\frac{\mu_3 v_L}{\sqrt{2}}\tan\zeta
\end{array}\right),~~~~~
\eea
where $\tan\zeta=k/v_R$, in analogy to $\tan\beta$ with left-right switch. These matrices can be diagonalized by the
unitary transformations: $V_1^\dag M_{1L}^+V_1=\text{diag}\left(M_{H_1^\pm},0\right)$ and $V_2^\dag M_{2R}^+V_2=\text{diag}\left(M_{H_2^\pm},0\right)$,
where
\be
\left(\begin{array}{c}\phi_1^{+} \\ \chi_L^{+}\end{array}\right)=
\underbrace{\left(\begin{array}{cc}
    \cos\beta & \sin\beta \\
    -\sin\beta & \cos\beta \\
  \end{array}\right)}_{V_1}
  \left(\begin{array}{c}H_1^+ \\ G_1^+\end{array}\right),\ \ \ \ \
\left(\begin{array}{c}\phi_2^+ \\ \chi_R^{+}\end{array}\right)=
\underbrace{\left(\begin{array}{cc}
    \cos\zeta & \sin\zeta \\
    -\sin\zeta & \cos\zeta \\
  \end{array}\right)}_{V_2}
  \left(\begin{array}{c}H_2^+ \\ G_2^+\end{array}\right).
\ee
The eigenstates $G_1^\pm$ and $G_2^\pm$ are the charged Goldstone
bosons eaten by the gauge bosons $W^\pm$ and $W'^\pm$ to acquire
their masses. The charged Higgs bosons masses are %
\bea
M_{H_1^\pm}^2&=&-(\al_2-\al_3)v_L^2\sec^2\beta-\sqrt{2} \mu_3 v_R\csc2\beta,\\
M_{H_2^\pm}^2&=&-(\al_2-\al_3)v_R^2\sec^2\zeta-\sqrt{2} \mu_3 v_L\csc2\zeta.
\eea
From these expressions, one can show that the mass of the charged Higgs can be of ${\cal O}(100)$ GeV.
%
\item \textbf{$CP$-odd Higgs bosons}:\\
We now turn to the neutral Higgs physical fields and their masses. This can be easily obtained if one develops the neutral components of the bidoublets $\Phi$ and the doublets $\chi_{L,R}$ around their vacua into real and imaginary parts, {\it i.e.,}
\be
\phi_i^0 = \frac{1}{\sqrt{2}} \left( v_i +\phi_i^{0R} +i \phi_i^{0I}\right), ~~~~~~ i = 1,2, L, R,
\ee
where $v_1=0,~v_2=k$, and $\phi_{L,R}=\chi_{L,R}$. In this case, the squared mass matrix of neutral Goldston and $CP$-odd Higgs bosons is given by%
\be
M_{I_{ij}}^2 = \frac{\partial^2 V(\Phi,\chi_{L,R})}{\partial \phi_i^{0I} \partial \phi_j^{0I}} \Big\vert_{\langle \phi_{i,j}^{0R}\rangle = \langle \phi_{i,j}^{0I}\rangle =0}.
\ee
One finds that this mass matrix in the basis of $(\phi_1^{0I}, \phi_2^{0I}, \chi_L^{0I}, \chi_R^{0I})$ is factored as a product of the squared mass of $\phi_1^{0I}$, which is totally decoupled due to the fact that we have $v_1=0$, times the following $3\times 3$ squared mass matrix of $(\phi_2^{0I}, \chi_L^{0I}, \chi_R^{0I})$:
\be
M_I^{2}=-\frac{k \mu_3}{2\sqrt{2}}
\left(\begin{array}{ccc}
\cot\beta\cot\zeta & -\cot\zeta & \cot\beta \\
-\cot\zeta & \tan\beta\cot\zeta & -1 \\
\cot\beta & -1 & \tan\zeta\cot\beta
\end{array}\right).
\ee
The mass of the first pseudoscalar Higgs boson $\phi_1^{0I} \equiv A_1$ is given by%
\be
M_{A_1}^2=2k^2\lm_2-(\al_2-\al_3)k\left(\cot^2\beta+\cot^2\zeta\right)-\frac{1}{\sqrt{2}}k \mu_3 \cot\beta\cot\zeta.
\ee
The matrix $M_I^2$ can be diagonalized by the unitary transformation: $U^\dag M^{2}_I U=\text{diag}\left(M_{A_2}^2,0,0\right)$,
\be
\left(\begin{array}{c}\phi_2^{0I} \\ \chi_L^{0I} \\ \chi_R^{0I}\end{array}\right)=
\underbrace{\left(\begin{array}{ccc}
 \frac{1}{\sqrt{\tan^2\beta+\tan^2\zeta+1}} & -\frac{\tan\zeta}{\sqrt{\tan^2\zeta+1}} & \frac{\tan\beta}{\sqrt{\left(\tan^2\zeta+1\right) \left(\tan^2\beta+\tan^2\zeta+1\right)}} \\
 -\frac{\tan\beta}{\sqrt{\tan^2\beta+\tan^2\zeta+1}} & 0 & \sqrt{\frac{\tan^2\zeta+1}{\tan^2\beta+\tan^2\zeta+1}} \\
 \frac{\tan\zeta}{\sqrt{\tan^2\beta+\tan^2\zeta+1}} & \frac{1}{\sqrt{\tan^2\zeta+1}} & \frac{\tan\beta \tan\zeta}{\sqrt{\left(\tan^2\zeta+1\right) \left(\tan^2\beta+\tan^2\zeta+1\right)}} \\
\end{array}\right)}_{U}
\left(\begin{array}{c}A_2 \\ G_1^0 \\ G_2^0\end{array}\right),
\ee
where $G_1^0$ and $G_2^0$ are the neutral Goldstone bosons eaten by the gauge bosons $Z$ and $Z'$ to acquire their masses. The other $CP$-odd Higgs bosons mass is given by
\bea
M_{A_2}^2=-\frac{k\mu_3}{\sqrt{2}}\frac{1+\tan^2\beta+\tan^2\zeta}{\tan\beta\tan\zeta}\label{Amass}.
\eea
It is worth mentioning that $M_{A_2}^2$ constrains the parameter $\mu_3$ to be negative. We find that the typical values of $CP$-odd Higgs masses are of ${\cal O}(100)$ GeV.

\item \textbf{$CP$-even Higgs bosons}:\\
Finally, we consider the $CP$-even Higgs bosons. Similar to the $CP$-odd Higgs, the squared mass matrix of $CP$-even Higgs bosons is given by%
\be
M_{R_{ij}}^2 = \frac{\partial^2 V(\Phi,\chi_{L,R})}{\partial \phi_i^{0R} \partial \phi_j^{0R}} \Big\vert_{\langle \phi_{i,j}^{0R}\rangle = \langle \phi_{i,j}^{0I}\rangle =0}.
\ee
Again, one finds that $H_1=\phi_1^{0R}$ is decoupled with mass $M_{H_1} = M_{A_1}$. The remaining squared mass matrix of the $CP$-even Higgs bosons is given in the basis $\left(\begin{array}{ccc}\phi_2^{0R} & \chi_L^{0R} & \chi_R^{0R}\end{array}\right)$ by
\be
M_R^2=\left(\begin{array}{ccc}
  k^2\lm_1-\frac{k\mu_3}{2\sqrt{2}}\cot\beta\cot\zeta & \al_{12} k^2\cot\beta+\frac{k\mu_3}{2\sqrt{2}}\cot\zeta & \al_{12} k^2\cot\zeta+\frac{k\mu_3}{2\sqrt{2}}\cot\beta\\
 \al_{12} k^2\cot\beta+\frac{k\mu_3}{2\sqrt{2}}\cot\zeta & k^2\lm_3 \cot^2\beta-\frac{k\mu_3}{2\sqrt{2}}\tan\beta\cot\zeta & k^2\lm_3 \cot\beta\cot\zeta-\frac{k\mu_3}{2\sqrt{2}} \\
 \al_{12} k^2\cot\zeta+\frac{k\mu_3}{2\sqrt{2}}\cot\beta & k^2\lm_3 \cot\beta\cot\zeta-\frac{k\mu_3}{2\sqrt{2}} & k^2\lm_3\cot^2\zeta-\frac{k\mu_3}{2\sqrt{2}}\tan\zeta\cot\beta
\end{array}\right).
\ee

This matrix can be diagonalized by a unitary transformation: $T^\dag M_R^2 T=\text{diag}\left(M_{H_2}^2,M_{H_3}^2,M_{H}^2\right)$.
The lightest eigenstate $H$ is the SM-like Higgs, the mass of which we will fix to be $125$ GeV. In general, from the numerical checks, we found that three $CP$-even Higgs
bosons ($H$ and $H_{1,3}$) are light [of ${\cal O}(100)$ GeV] and the other one $H_2$ is heavy [of ${\cal O}(1)$ TeV].

\end{enumerate}

\subsection{Couplings of the SM-like Higgs}
From the Yukawa Lagrangian (\ref{Yukawa}), one finds that the SM-like Higgs couplings with fermions in the ALRM are given by%
\bea
Y_{H\bar{u}u}&=&\frac{m_u}{v}\frac{T_{\Phi}}{\sin\beta},~~~~
Y_{H\bar{d}d}=\frac{m_d}{v}\frac{T_{L}}{\cos\beta},~~~~
Y_{H\bar{d}'d'}=\frac{m_{d'}}{v_R}T_{R},~~~~\\
Y_{H\bar{e}e}&=&\frac{m_e}{v}\frac{T_{\Phi}}{\sin\beta},~~~~
Y_{H\bar{n}n}=\frac{m_{n}}{v_R}T_{R},
\label{Fcoup}
\eea
where the elements $T_{\Phi},\ T_{L}$, and $T_{R}$ are the mixing couplings of the gauge eigenstates $\phi_2^{0R},\ \chi_L^{0R}$, and $\chi_R^{0R}$, respectively, with the lightest Higgs $H$.
Similarly, from the kinetic Lagrangian of the scalars, one can derive the following SM-like Higgs couplings with the electroweak gauge bosons,
\bea
g_{H W W}&=&g M_W \left(T_{\Phi}\sin\beta+T_{L}\cos\beta\right)\label{Wcoup},\\
g_{H W' W'}&=&g M_W\left(T_{\Phi}\sin\beta +T_{R}\frac{v_R}{v}\right),\\
g_{HZZ}&=&g_{LL}\cos^2\vartheta+g_{LR}\sin\vartheta\cos\vartheta+g_{RR}\sin^2\vartheta,\\
g_{HZ'Z'}&=&g_{LL}\sin^2\vartheta-g_{LR}\sin\vartheta\cos\vartheta+g_{RR}\cos^2\vartheta,
\eea
where
\bea
g_{LL}&=&\frac{g M_W}{\cos^2\tw}\left(T_{\Phi}\sin\beta+T_{L}\cos\beta\right),\\
g_{LR}&=&-\frac{\sqrt{2}g M_W}{\cos^2\tw\sqrt{\cos2\tw}}\left(T_{\Phi}\cos\beta\cos2\tw-T_{L}\sin\beta\sin^2\tw\right),\\
g_{RR}&=&\frac{g M_W}{\sqrt{2}\cos^2\tw\cos2\tw}
\left(T_{\Phi}\cos\beta\cos^22\tw +T_{L}\sin\beta\sin^4\tw+T_{R}\frac{v_R}{v}\cos^4\tw\right).
\eea
Finally, the SM-like Higgs couplings with the charged Higgs bosons are given by%
\bea
\lm_{HH_1^\pm H_1^\mp}&=& M_{1\Phi}T_{\Phi}+M_{1L}T_{L}+M_{1R}T_{R}\label{CH1coup},\\
\lm_{HH_2^\pm H_2^\mp}&=& M_{2\Phi}T_{\Phi}+M_{2L}T_{L}+M_{2R}T_{R}\label{CH2coup},
\eea%
where
\bea
M_{1\Phi}&=& 2 \left(k\lm_1\cos^2\beta-v_L(\al_2-\al_3)\cos\beta\sin\beta+k\al_{13}\sin^2\beta\right),\\
M_{1L}&=& 2 \left(v_L\al_{13}\cos^2\beta-k(\al_2-\al_3)\cos\beta\sin\beta+v_L\lm_3\sin^2\beta\right),\\
M_{1R}&=& 2 v_R\al_{12}\cos^2\beta-\sqrt{2}\mu_3\cos\beta\sin\beta+\sin^2\beta \left(2v_R\lm_3-\sqrt{2}\mu_3\tan\beta\right),\\
M_{2\Phi}&=& 2 \left(k\lm_1\cos^2\zeta-v_R(\al_2-\al_3)\cos\zeta\sin\zeta+k\al_{13}\sin^2\zeta\right),\\
M_{2L}&=& 2 v_L\al_{12}\cos^2\zeta-\sqrt{2}\mu_3\cos\zeta\sin\zeta+\sin^2\zeta \left(2v_L\lm_3-\sqrt{2}\mu_3\tan\zeta\right),\\
M_{2R}&=& 2 \left(v_R\al_{13}\cos^2\zeta-k(\al_2-\al_3)\cos\zeta\sin\zeta+v_R\lm_3\sin^2\zeta\right).
\eea
\section{ALRM effects in $H \to \gamma \gamma$ decay}
As advocated in the Introduction, CMS and ATLAS collaborations
observed a SM-like Higgs boson with mass around $125$ GeV and
signal decay strengths as given in Eqs. (\ref{CMS})-(\ref{ATLAS}).
For instance, CMS found \cite{Khachatryan:2014ira,Moriond1-CMS:2013,CMSNOTE1:2013}%
\bea
\mu_{\gamma \gamma}=\mu(H \to \gamma \gamma) &=& 1.14^{+0.26}_{-0.23}, \label{CMS}\\
\mu_{ZZ}=\mu(H \to Z Z) &=& 0.91^{+0.3}_{-0.24}, \\
\mu_{WW}=\mu(H \to W W) &=& 0.76 \pm 0.21, %
\eea
while the ATLAS experiment reported that the signal strength of these decays are given by \cite{Aad:2014eha,Moriond2-ATLAS:2013,ATLASNOTE2:2013}: %
\bea
\mu_{\gamma \gamma}=\mu(H \to \gamma \gamma) &=& 1.17 \pm 0.27, \\
\mu_{ZZ}=\mu(H \to Z Z) &=& 1.7 \pm 0.5, \\
\mu_{WW}=\mu(H \to W W) &=& 1.01 \pm 0.31.\label{ATLAS}%
\eea
These results indicate enhancement in the diphoton
decay channel, with more than $2\sigma$ deviation, which could be
a very important signal for possible new physics beyond the SM.
Much work has been done to accommodate these results in different
extensions of the SM \cite{Carena:2012xa,Chao:2012xt,Picek:2012ei,Chang:2012ta,Dev:2013ff,Basak:2013eba,Cao:2013wqa,Feng:2013mea,Berg:2012cg,Khalil2013473}.
The Higgs signal strength of the decay channel, $H\to \gamma\gamma$, relative to the SM
expectation is defined as
\bea
\mu_{\gamma\gamma} &=& \frac{\sigma(pp \to H\to \gamma\gamma)}{\sigma(pp \to H \to
\gamma\gamma)^{\text{SM}}}= \frac{\sigma(pp \to H)}{\sigma(pp \to H)^{\text{SM}}} ~\frac{\text{BR}(H\to \gamma\gamma)}{\text{BR}(H\to \gamma\gamma)^{\text{SM}}} \nonumber\\
&=& \frac{\Gamma(H\to g g)}{\Gamma(H\to gg)^{\text{SM}}}
~\frac{\Gamma_{\text{tot}}^{\text{SM}}}{\Gamma_{\text{tot}}} ~\frac{\Gamma(H\to A
A)}{\Gamma(H\to \gamma\gamma)^{\text{SM}}} = \kappa_{gg} \cdot \kappa_{\text{tot}}^{-1} \cdot \kappa_{\gamma\gamma}~,%
\eea
where $\sigma(pp \to H)$ is the total Higgs production cross
section and $\text{BR}(H\to \gamma\gamma)$ is the branching ratio of the
corresponding channel. The total Higgs decay width is given by the
sum of the dominant Higgs partial decay widths,
$\Gamma_{\text{tot}} = \Gamma_{b\bar{b}} + \Gamma_{WW} + \Gamma_{ZZ} +
\Gamma_{gg} + \Gamma_{\tau\bar{\tau}}$. Other partial decay widths are much
smaller and can be safely neglected. In the SM with 125 GeV Higgs
mass, these partial decay widths are given by:
$\Gamma_{b\bar{b}} = 2.3\times 10^{-3}$ GeV, $\Gamma_{WW}= 8.7
\times 10^{-4}$ GeV, $\Gamma_{ZZ}= 1.1 \times 10^{-4}$ GeV, $\Gamma_{gg}=3.5 \times 10^{-4}$ and
$\Gamma_{\tau\bar{\tau}}=2.6 \times 10^{-4}$ GeV.  As shown in
the previous section, the Higgs couplings $g_{HWW}$ and $Y_{Hb\bar{b}}$ may slightly change from the SM values.
Hence, the total decay width of the Higgs boson remains very close to the SM result.
This has been confirmed numerically, and to a very good
approximation, one can consider $\kappa_{\text{tot}}\simeq1$.

\begin{figure}[t]
\begin{center}
\includegraphics[scale=0.6]{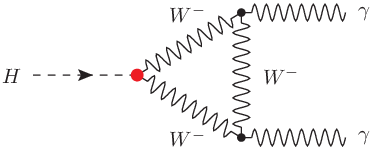}~~~~~~\includegraphics[scale=0.55]{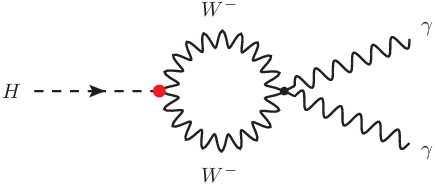}~~~~~\includegraphics[scale=0.6]{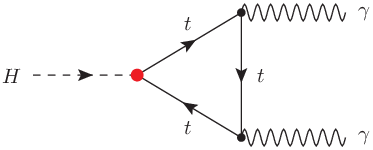}
\\$~$\\
\includegraphics[scale=0.6]{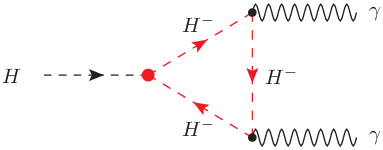}~~~~~~~~~~~\includegraphics[scale=0.55]{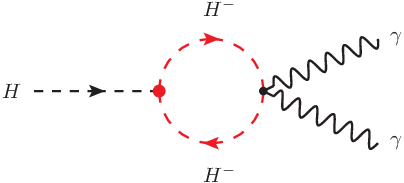}
\caption{\label{feynman} Feynman diagrams for the Higgs decay $H\to\gamma\gamma$ mediated by gauge bosons $W^\pm$, a top quark, and charged scalars $H^\pm$.}
\end{center}
\end{figure}

Now, we turn to the SM-like Higgs decay into a diphoton, $W^+ W^-$
and $Z Z$ in our ALRM. As shown in the previous section, the low-energy effective theory of the ALRM contains two charged Higgs bosons
that can be light, of ${\cal O}(100)$ GeV, and may give
relevant contributions to the SM-like Higgs decay into a diphoton.
In addition, the couplings of the SM-like Higgs with a top quark and
$W$ gauge boson may be suppressed or even flipped, which would lead to
significant enhancement/suppression in $\Gamma(H\to \gamma
\gamma)$. The Feynman diagrams of the Higgs decay $H \to \gamma
\gamma$, mediated by the gauge bosons $W^\pm$, top quark, and
light-charged Higgs bosons are shown in Fig. \ref{feynman}. Note
that in the conventional LRM there are interaction vertices among
charged Higgs, the $W$-boson and neutral Higgs/photon; therefore, another four diagrams with $W^\pm$ and $H^\pm$ running in the loop of
triangle diagrams can be generated. In our ALRM, these vertices
identically vanish due to the discrete $S$ symmetry. In
this case, the one-loop partial decay width of the $H$ decay into
two photons is given by \cite{Carena:2012xa}
\bea
\Gamma\left(H \to \gamma \gamma\right) =\frac{\al^2 m_H^3}{1024 \pi^3}
\Big|\frac{g_{HWW}}{M_W^2} Q_W^2 F_1(x_W) + N_{c,t} Q_t^2
\frac{ 2 Y_{H\bar{t}t}}{m_t} F_{1/2}(x_t)  + \sum_{i=1}^2
Q_{H_i^\pm}^2\frac{\lm_{H H_i^\pm H_i^\mp}}{M_{H_i^\pm}^2} F_0(x_{H_i^\pm})
\Big|^2,
\eea
where $x_t = M_H^2/4m_t^2,\ x_k = M_H^2/4M_k^2$, $k=W,H_{1,2}^\pm$. The color factor and
electric charges are given by $N_{c,t}=3$, $Q_{W}=$~$Q_{H_i^+}=1$, and
$Q_t=2/3$. Recall that the relevant Higgs couplings in the ALRM are
given by $g_{HWW},\ Y_{H\bar{t}t}$, and $\lm_{H H_i^\pm H_i^\mp}$ in Eqs. (\ref{Fcoup}), (\ref{Wcoup}), (\ref{CH1coup}), and (\ref{CH2coup}),
with $T_{\Phi} \sim T_{L} \gg T_{R}$. Finally, the loop functions $F_i(x)$ are given by \cite{Carena:2012xa}%
\bea
F_1(x) &=& -\left[2 x^2 + 3 x + 3 (2x -1) \arcsin^2(\sqrt{x})\right] x^{-2},\\
F_{1/2}(x) &=& 2 \left[x +(x-1) \arcsin^2(\sqrt{x})\right] x^{-2},\\
F_{0}(x) &=& - \left[x -\arcsin^2(\sqrt{x})\right] x^{-2}.%
\eea

\begin{figure}[th]
\includegraphics[scale=0.55]{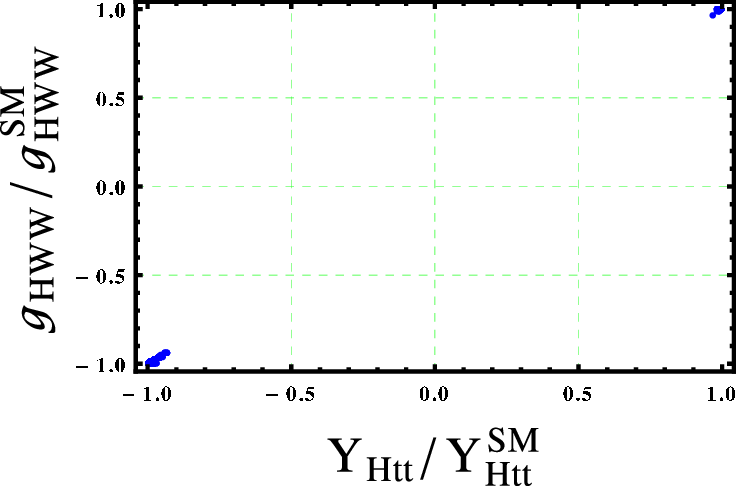}~~~~~~~~~~\includegraphics[scale=0.55]{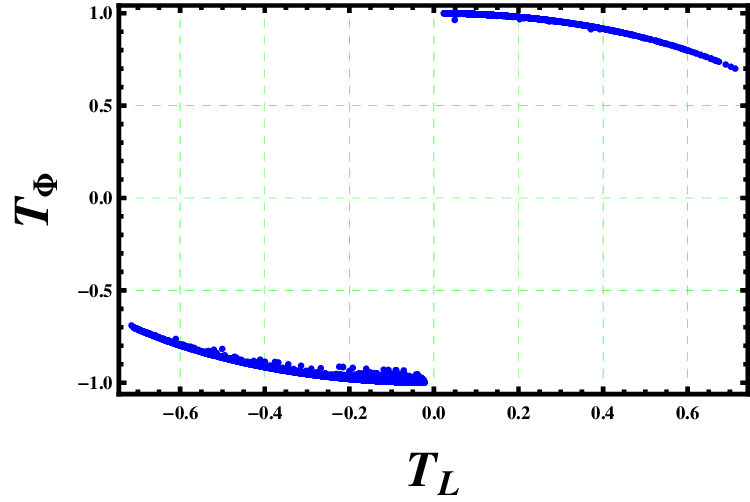}
\caption{(Left panel) The relation between the coupling ratios $g_{HWW}/g_{HWW}^{\text{SM}}$ and $Y_{H\bar{t}t}/Y_{H\bar{t}t}^{\text{SM}}$. (Right panel) The relation between the mixing parameters $T_\Phi$ and $T_L$.}
    \label{wt-ratios}
\end{figure}

For Higgs mass of order $125$ GeV and charged Higgs mass of order
200 GeV, the loop functions $F_1(x_W)$, $F_{1/2}(x_t)$, and
$F_0(x_{H^\pm})$ are of order $-8.32$, $+1.38$, and $+0.43$,
respectively. Therefore, the partial decay width $\Gamma(H\to
\gamma \gamma)$ can be enhanced through one of the following
possibilities: $(i)$ large charged Higgs couplings such that
$\lm_{H H^\pm H^\mp}/M_{H^\pm}^2$ is of order $g_{HWW}/M_{W}^2$, and with an opposite sign to compensate the difference
in sign between $F_1(x_W)$ and $F_0(x_{H^\pm})$; $(ii)$ either the sign of the top
Yukawa coupling, $Y_{H\bar{t}t}$, or the sign of the coupling between the $W$ boson and the SM-like Higgs, $g_{HWW}$, is flipped so that a constructive interference
between $W$-gauge boson and top-quark contributions takes place;
and $(iii)$ a significant reduction for the top Yukawa coupling, $Y_{H\bar{t}t}$, to
minimize the destructive interference between $W$ and $t$
contributions. In Fig. \ref{wt-ratios}, we display the changes in
$g_{HWW}$ and $Y_{H\bar{t}t}$, normalized to their SM values.
As can be seen from this figure, both couplings are slightly changed from their
expectations in the SM. In addition, both $g_{HWW}$ and $Y_{H\bar{t}t}$ may flip their sign simultaneously, and hence the usual destructive  interference
between $W$-gauge boson and top-quark contributions remains intact. Therefore,  one would not expect any enhancement of $\Gamma(H\to
\gamma \gamma)$. The sign correlation between the coupling ratios can be understood from the fact that the parameters
$T_{\Phi}$ and $T_{L}$ in Eqs. (\ref{Fcoup}) and (\ref{Wcoup}), which lead to
the modifications of these couplings, have the same sign in the
allowed region of ALRM parameter space, as shown in Fig. \ref{wt-ratios}.

The Higgs boson production at the LHC is dominated by gluon-gluon
fusion. As in the SM, this channel is mediated by top quarks via a
one-loop triangle diagram. The extra quark $d'$ gives a negligible
contribution to $\sigma(gg \to H)$ due to the suppression of its coupling
with the SM-like Higgs and also its large mass. As mentioned, the top
Yukawa coupling can be slightly different from the SM coupling;
therefore, the ratio $\kappa_{gg}=\Gamma(H\to~gg)/\Gamma(H\to~gg)^{\text{SM}}$ can be slightly deviated from $1$.

\begin{figure}[t]
\begin{center}
    \includegraphics[scale=0.55]{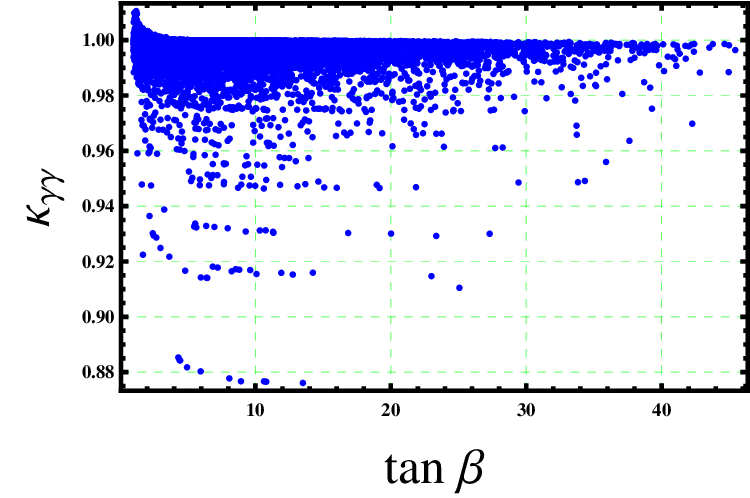}~~~~~~~\includegraphics[scale=0.78]{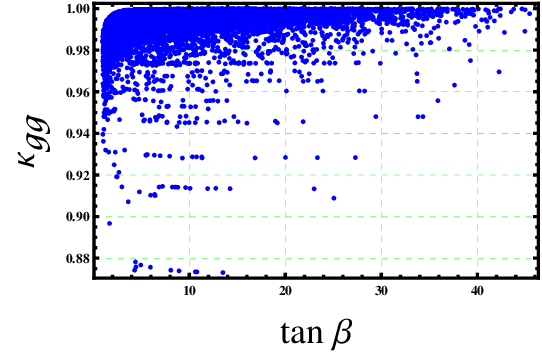}
    \caption{$\kappa_{\gamma\gamma}$ and $\kappa_{gg}$ as functions of $\tan\beta$ and the parameters $\lm_3,~\al_1,~\al_2$ and $M_{H^\pm_{1,2}}$.}
    \label{kAA}
\end{center}
\end{figure}

In Fig. \ref{kAA}, we display the results of $\kappa_{\gamma \gamma} =
\Gamma(H\to \gamma \gamma)/\Gamma(H\to \gamma \gamma)^{\text{SM}}$ and
$\kappa_{gg}$ as function of $\tan \beta$ for
$0<\lm_1,\lm_3,\lm_4<\sqrt{4\pi},~-\sqrt{4\pi}<\lm_2<0,~-\sqrt{4\pi}<\al_1,\al_2,\al_3<\sqrt{4\pi},~100<M_{H^\pm_{1,2}}<300$, and $\mu_3<0$,
to be consistent with the perturbative unitarity and the minimization and boundedness from below conditions (\ref{stcond})$-$(\ref{mcc}). It is worth mentioning that
for $\mu_3<0$, one finds, from the minimization conditions, that
$\lm_4-\lm_3>0$, and from (\ref{stcond}), $\lm_3\geq0$ and hence $\lm_4>\lm_3\geq0$.
In our numerical analysis, we express the parameters
$\mu_1^2,\ \mu_2^2$, and $\lm_4$ in terms of the three vevs
$v_L,\ v_R$, and $k$ (or $v$, $\tan\beta$, and $M_{W'}$). We also
substitute the parameters $\mu_3$ and $\al_3$ in terms of the charged Higgs
masses $M_{H_{1,2}^\pm}$ and the parameter $\lm_1$ in terms of the SM-like Higgs
mass $M_H=125$ GeV. Thus, one can write the matrix
$T\equiv
T\left(\tan\beta,M_{W'},M_{H_{1,2}^\pm},\lm_3,\al_1,\al_2\right)$. This
figure  confirms our theoretical expectation and shows that both of  $\kappa_{\gamma\gamma}$ and
$\kappa_{gg}$ can slightly deviate from $1$.

\begin{figure}[t]
\begin{center}
    \includegraphics[scale=0.55]{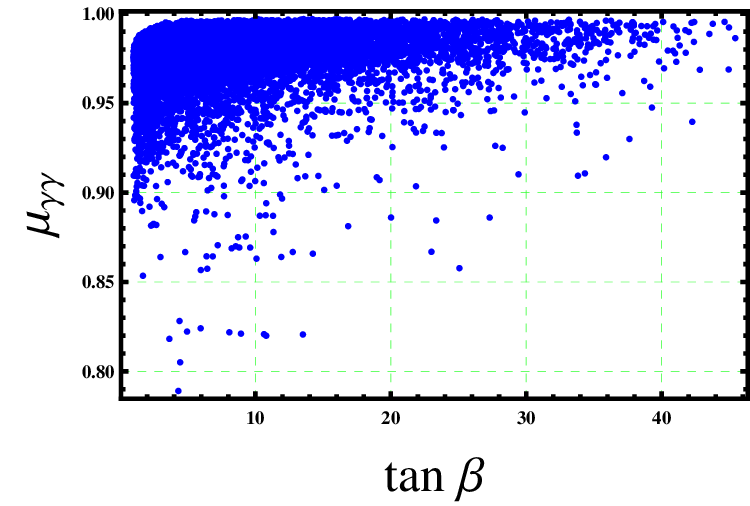}~~~~~~~~~~~\includegraphics[scale=0.55]{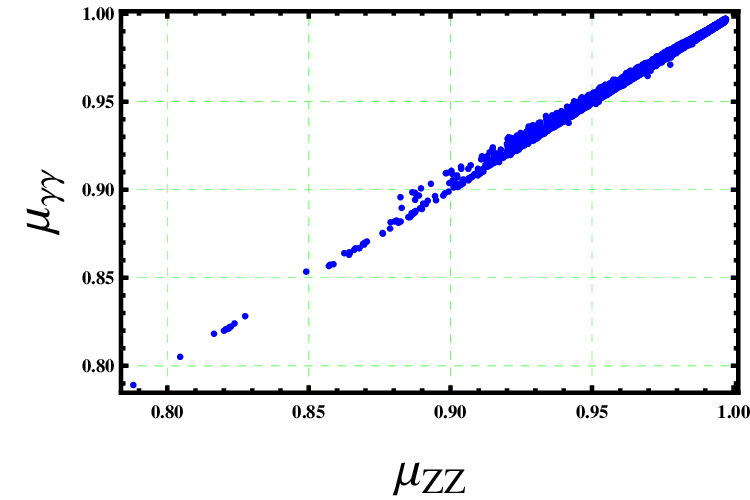}
    \caption{(Left panel) The signal strength $\mu_{\gamma\gamma}$ as a function of $\tan\beta$ and the parameters $\lm_3,~\al_1,~\al_2$ and $M_{H^\pm_{1,2}}$. (Right panel) Correlation between
    $\mu_{\gamma \gamma}$ and $\mu_{ZZ}$ in the ALRM.}
    \label{Rgg-RZZ}
\end{center}
\end{figure}

In this case, it is clear that the signal strength
$\mu_{\gamma \gamma}$ is also close to the SM expectation and can be still consistent with  both ATLAS and
CMS experimental results. In Fig. \ref{Rgg-RZZ}, we show the signal
strength as a function of $\tan \beta$, where other parameters are scanned in the above-mentioned regions. For completeness, we also present the
correlation between $\mu_{\gamma \gamma}$ and $\mu_{ZZ}$, which
equals $\mu_{WW}$ in our model. It is remarkable that all signal strengths of Higgs decay channels in the ALRM are slightly less than the SM results.

\section{Signatures at the LHC}
In this section we study the interesting signatures of the exotic
quark $d'$ associated with our ALRM at the LHC. In particular, we
will analyze and compute the cross section for the production of
this heavy quark and its subsequent decays into jets, leptons and
missing energy. The Lagrangian of $d'$ interactions with the SM
quarks can be derived
from (\ref{Yukawa}) as %
\bea%
\mathcal{L}_\text{Y}^{d'} &=& -\overline{u}~\left(\cos\zeta Y^q P_R+\sin\zeta Y^q_R P_L\right)~H_2^+ V'_{\text{CKM}}~d'+\text{h.c.}~,\label{DC}%
\eea%
where $V'_{\text{CKM}}$ is the right-handed quark mixing matrix. In addition,
the kinetic Lagrangian of $d'$ leads to the following interactions
with the gauge bosons
\bea
\mathcal{L}_{\text{gauge}}^{d'} &=& -\frac{i g_s}{2}\overline{d}'\gamma^\mu \lm_a G_\mu^a d'
-\frac{i g}{\sqrt{2}}\overline{u} P_R \gamma^\mu {W'}_\mu^+ V'_{\text{CKM}} d'
-\frac{i g}{\sqrt{2}}\overline{d}'\gamma^\mu P_R {W'}_\mu^-{V'}_{\text{CKM}}^\dag u\nn
\\&&
+\frac{i}{3}e\overline{d}'\gamma^\mu
\left[A_\mu+\left(\hat{P}\sin\vartheta-\frac{1}{2}\tan\theta_w\cos\vartheta\right)Z_\mu +\left(\hat{P}\cos\vartheta+\frac{1}{2}\tan\theta_w\sin\vartheta\right)Z'_\mu \right] d',
\eea
where
\be
\hat{P}=\frac{3\cos2\theta_w-\sin^2\theta_w}{6\sin\theta_w\cos\theta_w\sqrt{\cos2\theta_w}}P_R
+\frac{\sin\theta_w}{\cos\theta_w\sqrt{\cos2\theta_w}}P_L,
\ee
where $\lm_a$'s,
$a=1,...,8$, are the Gell$-$Mann matrices and $\vartheta$ is given
in (\ref{Z-Z' Mixing}). Accordingly, in this case the pair production of $d'$ at the LHC is dominated by the following channel:
$gg\rightarrow d' \bar{d'}$. Considering all contributions from s,t, and u-channels, the squared amplitude of this process is given by%
\bea
\Big|{\cal M}(gg\rightarrow d' \bar{d'})\Big|^2&=&\frac{g_s^4}{24\hat{s}^2}
\frac{\left(9 m_{d'}^4-9 m_{d'}^2 (\hat{s}+2 \hat{t})+4 \hat{s}^2+9 \hat{s} \hat{t}+9\hat{t}^2\right)}{\left(m_{d'}^2-\hat{t}\right)^2}\nn
\\&\times&\Big[\frac{m_{d'}^2 \left(\hat{s}^3+2 \hat{s}^2 \hat{t}+8 \hat{s} \hat{t}^2+8 \hat{t}^3\right)+\hat{t} (\hat{s}+\hat{t}) \left(\hat{s}^2+2 \hat{s} \hat{t}+2 \hat{t}^2\right)}{\left(-m_{d'}^2+\hat{s}+\hat{t}\right)^2}\nn\\
&-&\frac{2 m_{d'}^8-8 m_{d'}^6 \hat{t}+m_{d'}^4 \left(3 \hat{s}^2+4 \hat{s} \hat{t}+12 \hat{t}^2\right)}{\left(-m_{d'}^2+\hat{s}+\hat{t}\right)^2}\Big].~~~~~
\eea
In addition the squared amplitude of the pair production of $d'$ through the channel $q \bar{q} \to \gamma/g \to d' \bar{d'}$ is given by
\bea
\left|{\cal M}\left(q\bar{q}\rightarrow \gamma/g\rightarrow d'\bar{d}'\right)\right|^2&=&\frac{4 \left(2 g^4+9 g_s^4\right)}{81 \hat{s}^2}
\left(2\hat{s}\hat{t}-4\left(m_q^2+m_{d'}^2\right)\hat{t}+2\left(m_q^2+m_{d'}^2\right)^2+2\hat{t}^2+\hat{s}^2\right),
\eea
where $\hat{s}, \hat{t}$ are the partonic Mandelstam variables.
The differential cross section is given by
\be
\frac{d\hat{\sigma}}{d\cos\theta}=\frac{B}{16\pi\hat{s}^2}|{\cal M}|^2,
\ee
where
$B =\sqrt{1- \left(4m_{d'}^2/\hat{s}\right)}$.
The cross section of  $pp \to d' d'$ is given by
\be
\frac{d\sigma}{d\cos\theta}=\sum_{i,j}~ \int_{x_0}^1 dx f_{i}(x) f_{j} \Big(\frac{4 m_{d'}^2}{s x}\Big)~\frac{d\hat{\sigma}}{d\cos\theta},
\ee
where $i,j$ refer to the partons. The partons energy fractions are
given by $x_1 x_2 = \hat{s}/s,$ so that the minimum parton energy
fraction to produce the $d' d'$ pair is $x_0=2m_{d'}/\sqrt{s}$. Also, $\hat{t} =-\frac{1}{2}\hat{s}\left(1-B\cos\theta\right)+M^2_{d'}$.
Therefore, one finds that the production cross
section is given by%
\be
\frac{d\hat\sigma}{d\hat{t}}=\frac{1}{8\pi\hat{s}^3}\left|{\cal M}\right|^2.
\ee

\begin{figure}[t]
\begin{centering}
    \includegraphics[scale=0.6,angle=0]{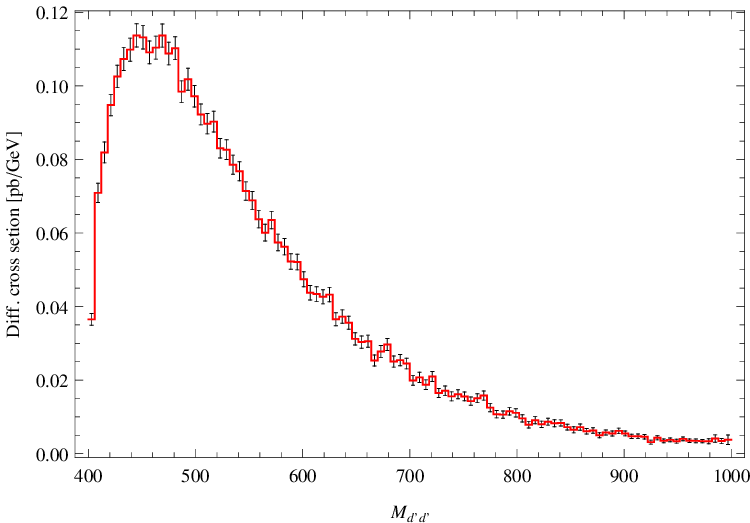}~~\includegraphics[scale=0.6,angle=0]{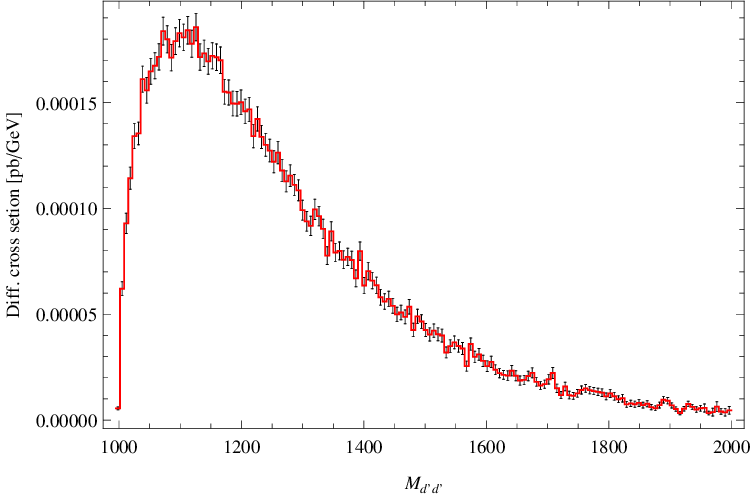}
    \caption{\label{shat} Differential production cross section of exotic quark $d'$ as a function of the invariant mass $M_{d'd'}$. In the left panel, $m_{d'}=300$ GeV, and in the right panel, $m_{d'}=500$ GeV.}
    \label{sigma}
\end{centering}
\end{figure}

In Fig. \ref{sigma}, we display the differential cross section of the $d'$ pair
production at the LHC with $\sqrt{s} = 14$ TeV as a function of the
invariant mass $M_{d'd'}$ for two choices of $m_{d'}$, namely, $m_{d'} =300$ and $500$ GeV.
As can be seen from this figure, the typical value of the $d'$ production cross section is of ${\cal O}(1)$ fb, which was quite
accessible at the LHC during its second run. The dominant decay channel of the produced $d'$ quark is given by $d' \to H_2^+ u$, as indicated in (\ref{DC}). One can show that the corresponding decay rate is given by %
\be %
\Gamma\left(d'\rightarrow u H_2^+\right)=\frac{|V'_{\text{CKM}}|^2}{16\pi\hbar}\left(|Y^q|^2\cos^2\zeta+|Y^q_R|^2\sin^2\zeta\right) m_{d'}\left(1-\frac{M_{H_2^+}^2}{m_{d'}^2}\right)^2. %
\ee %
Here, we assumed that $m_u\ll m_{d'}$. On the other hand, the
charged Higgs boson $H_2^+$ decays into a lepton and scotino through
the interactions%
\be
\mathcal{L}_\text{Y}^{H_2^+}=\overline{n}~H_2^+~U'_{\text{MNS}}\left(\cos\zeta Y^\ell P_L + \sin\zeta Y^\ell_R P_R\right)~e+\text{h.c.}~.
\ee %
Thus, the decay rate of $H_2^- \to e^-n$, for {\bf $m_e=0$}, is given by
\be
\Gamma\left(H_2^-\rightarrow e^-n\right)=\frac{|U'_{\text{MNS}}|^2}{16\pi\hbar}\left(|Y^\ell|^2\cos^2\zeta+|Y^\ell_R|^2\sin^2\zeta\right) M_{H_2^+} \left(1-\frac{m_n^2}{M_{H_2^+}^2}\right)^2.
\ee
\begin{figure}[t]
\begin{centering}
    \includegraphics[scale=0.8,angle=0]{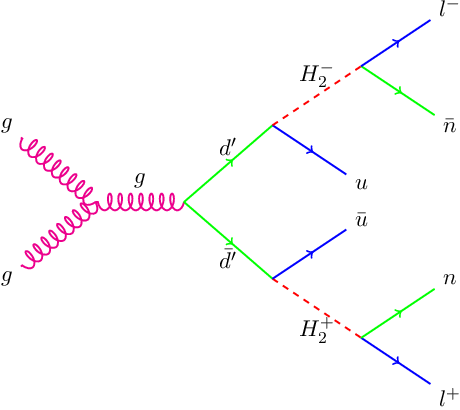}
    \caption{The exotic quark, $d'$, creation and decay.}
    \label{eq}
\end{centering}
\end{figure}

\begin{figure}[t]
\begin{centering}
\includegraphics[scale=0.5,angle=0]{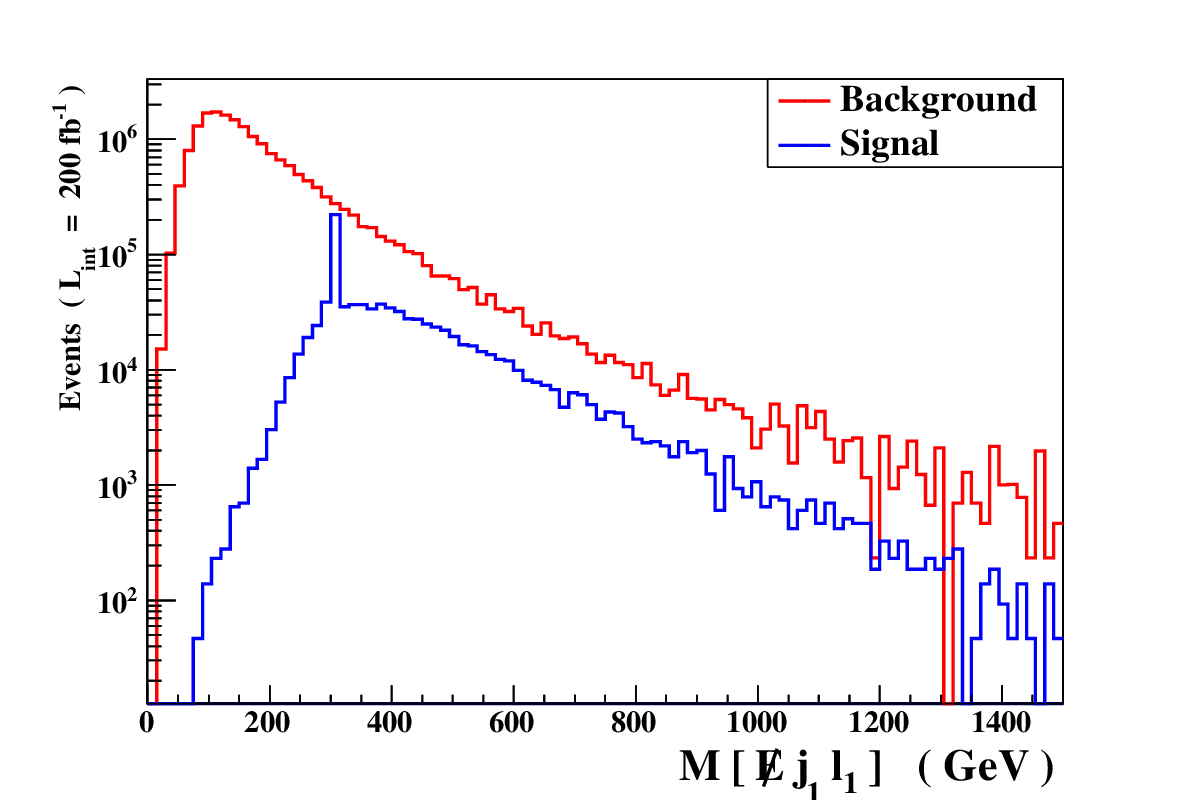}
\caption{The reconstructed invariant mass of extra quark, $d'$, which decays to $l+ jet+ {\rm missing~ energy}$ and its background for $m_{d'}=300$ GeV. No cut has been imposed yet.}
\label{no-cut}
\end{centering}
\end{figure}

In Fig. \ref{eq}, we show the total cross section of this process with
an opposite-sign dilepton, which is the most striking
signature for this exotic quark at the LHC. This cross section can be approximately written as %
\bea \sigma\left(g g\rightarrow g\rightarrow
d'\overline{d}'\rightarrow l^\mp
l^\pm+E_T^{\text{miss}}+\text{jets}\right)&\simeq& \sigma\left(gg
\rightarrow g\rightarrow d'\overline{d}'\right)\nn
\\&\times&\text{BR}\left(d'\rightarrow H_2^+ +\text{jets}\right)^2
\text{BR}\left(H_2^\mp \rightarrow l^\mp + E_T^{\text{miss}}\right)^2 .%
\eea%
Since the dominant decay channel of $d'$ is $d'\rightarrow u H_2^-$ and
the charged Higgs decays mainly to $l^\pm + n$, one finds $\text{BR}\left(d'\rightarrow u H_2^-\right)\simeq1$ and
$\text{BR}\left(H_2^\pm\rightarrow l^\pm n\right)\simeq1$. Therefore, $\sigma\left(g g\rightarrow g\rightarrow
d'\overline{d}'\rightarrow l^\mp
l^\pm+E_T^{\text{miss}}+\text{jets}\right) \simeq \sigma\left(g g\rightarrow g\rightarrow
d'\overline{d}'\right) \simeq {\cal O}(1)$ fb, which can be accessible at the LHC with $\sqrt{s}\simeq 14$ TeV.
In Fig. \ref{no-cut}, we show the reconstructed invariant mass of the extra quark $d'$, which decays into $l+ n ({\rm scotino}) + jet$, with all possible
background. In this figure, we have not imposed any cut yet. Therefore, the background is clearly dominated the signal. Here, we assume $m_{d'}=300$ GeV, the charged Higgs mass is of order $200$ GeV, and the LHC integrated luminosity is of order $200$ fb$^{-1}$.

\begin{figure}[t]
\begin{centering}
\includegraphics[scale=0.39,angle=0]{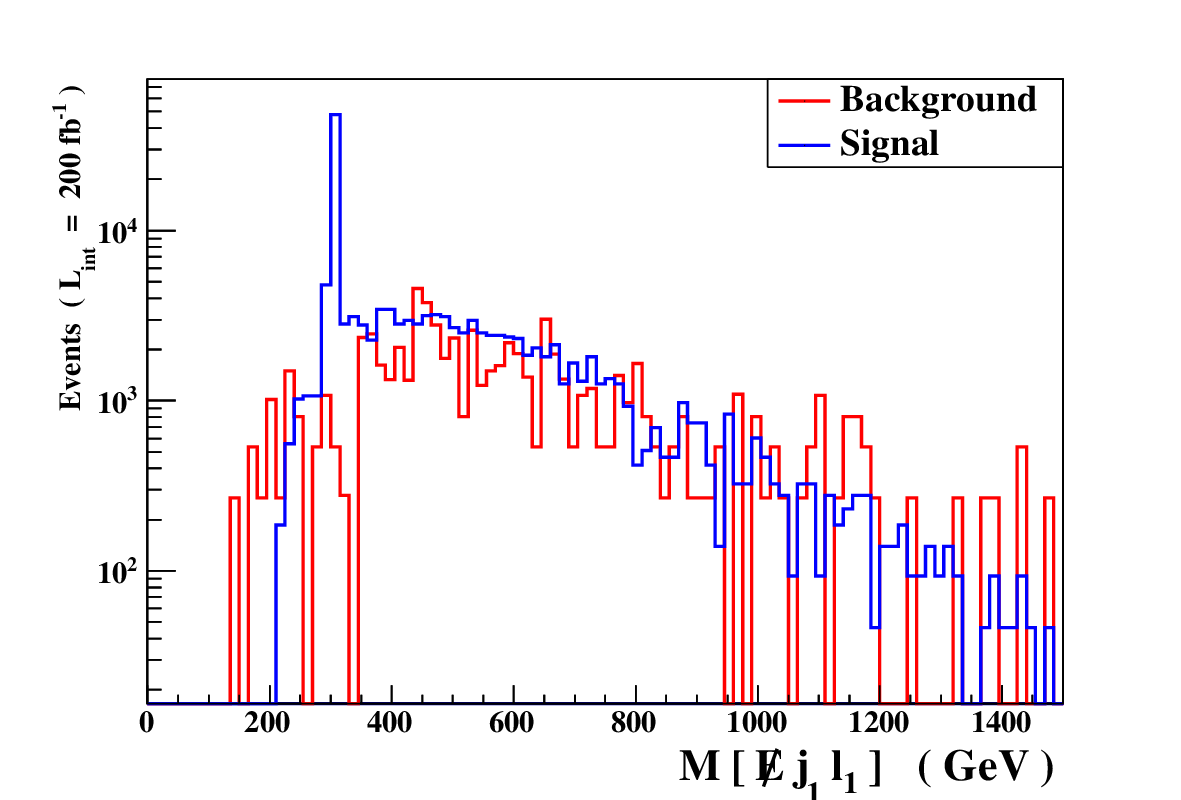}~\includegraphics[scale=0.39,angle=0]{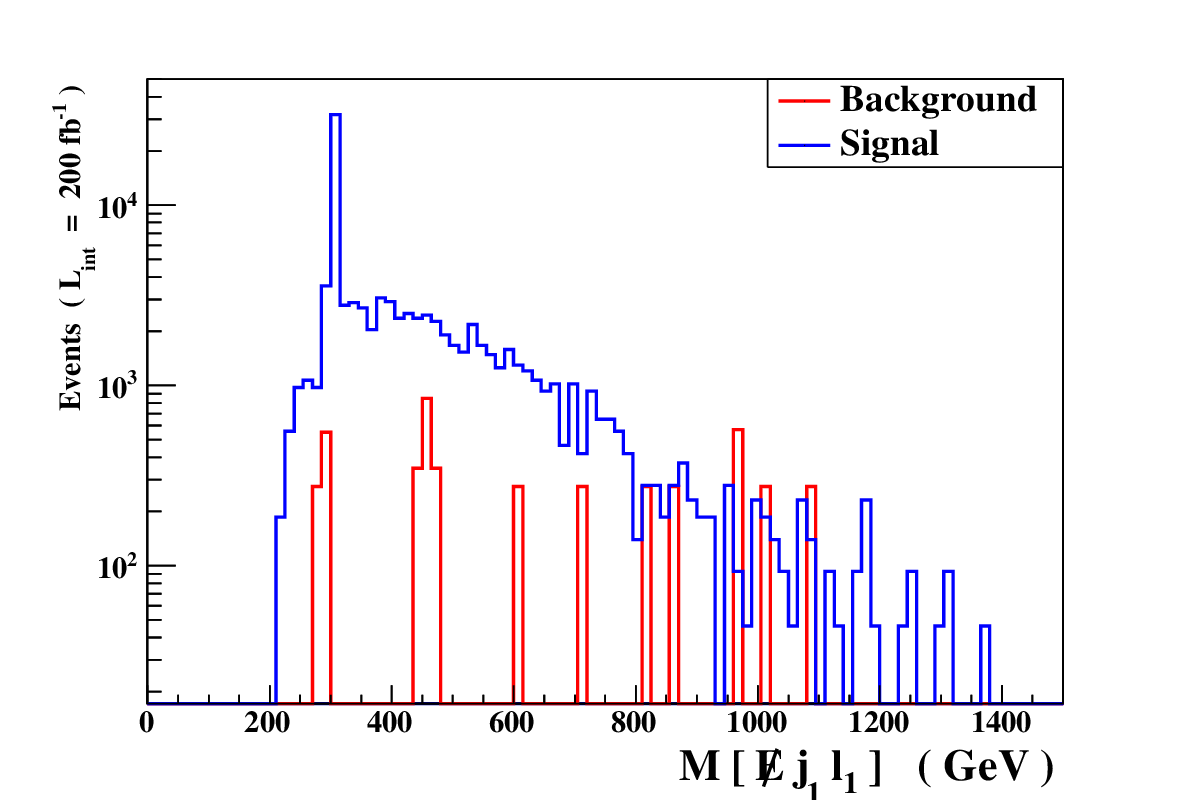}
\caption{The reconstructed invariant mass of the extra quark, $d'$, which decays to $l+ jet+ {\rm missing~ energy}$ for $m_{d'}=300$ GeV, with $\slashed{E}_T>200$ GeV cut (left panel) and $H_T < 200$ GeV cut (right panel).}
    \label{with-cut}
\end{centering}
\end{figure}

In Fig. \ref{with-cut} (left panel), we plot the number of reconstructed events per bin of the invariant mass of $d'$ of the above process for signal and SM background at $\slashed{E}_T~{\rm cut} > 200$ GeV, where $\slashed{E}_T$ is the missing transverse energy, $\slashed{E}_T=\left|\left|\sum_{\text{visible particles}}\vec{p}_T\right|\right|$, with $m_{d'}=300$ GeV and $\sqrt{s}=14$ TeV. This figure shows that it is possible to extract a good significance for the extra-quark signal in this channel. In addition, we also impose a cut, $H_T <200$ GeV, where
$H_T$ is the total transverse hadronic energy: $H_T=\sum_{\text{hadronic particles}}\left|\left|\vec{p}_T\right|\right|$. It is remarkable that with $H_T$ cuts the signal can be much larger than the background.
We have used Feynrules \cite{Alloul:2013bka} to generate the model files and Calchep \cite{Belyaev:2012qa} and MadEvent5 \cite{Alwall:2011uj,Conte:2012fm}
to calculate the numerical values of the cross sections and number of events, respectively.

Finally, we provide in Table \ref{SB} some details for the used cuts on $P_T$ and $H_T$ on the signal and background for the process $pp \to d' d' \to (l^- l^+)+ (u\bar{u})+(n n) $.
As can be seen from the results in this table, the signal of this process can be much larger than the background if one imposes the proper $H_T$ cuts. It is worth mentioning that the Higgs sector of our model is very similar to the two Higgs doublets in the minimal supersymmetric standard model, where one Higgs doublet couples to up quarks and the second couples to down quarks. Therefore, it does not lead to any flavor-changing neutral current problem, and a light charged Higgs is phenomenologically acceptable. The number of events of exotic quark $d'$ at the LHC may slightly changed if a heavier charged Higgs is considered, but with keeping $m_{H^+} < m_{d'}$, to ensure that $\text{BR}(d’ \rightarrow H^+ + jets) \sim 1$.

\begin{table}[!ht]
\centering
\begin{tabular}{|c|c|c|c|}
\hline
Cuts (GeV)   & Signal (S) & Background (B) & S vs B \\ [0.5ex]\hline
Initial (no cut) & 463999 & 9309732 $\pm$ 21646 & 0.049840 $\pm$ 0.000116 \\ [0.5ex]\hline
Cut 1 ($\slashed{E}_T>200$) & 72291 $\pm$ 247 & 33523 $\pm$ 198 & 2.1564 $\pm$ 0.0148 \\ [0.5ex]\hline
Cut 2 ($H_T<200$) & 47977 $\pm$ 207 & 1942.7 $\pm$ 44.3 & 24.696 $\pm$ 0.573 \\ [0.5ex]\hline
\end{tabular}
\caption{Signal vs background for the process $pp \to d' d' \to (l^- l^+)+ (u\bar{u})+(n n) $ with/without cuts.}
\label{SB}\end{table}

\section{Conclusions}

In this paper, we have analyzed some phenomenological aspects of the
alternative left-right model, motivated by the superstring-inspired $E_6$ model.
We provided a detailed analysis for the symmetry breaking and Higgs sector of this model,
which consists of four neutral $CP$-even Higgs, two $CP$-odd Higgs and two charged Higgs bosons. We emphasized that
three neutral $CP$-even Higgs and two $CP$-odd Higgs in addition to two charged Higgs can be light, of
${\cal O}(100)$ GeV. We also found that the contributions of charged Higgs bosons and the extra exotic quark $d'$
to $H \to \gamma \gamma$ are quite negligible. Therefore, our model predicts signal
strengths of Higgs decay, in particular, of $H \to \gamma \gamma$ and $H \to W^+ W^-$ that coincide with the SM expectations.

Finally, we studied the striking signatures of the exotic down-type quark at the LHC.
In particular, we computed the cross section of $d'$-pair production .
We showed that the typical value of this cross section is of ${\cal O}(1)$ fb, which is quite accessible at the LHC.
The decay of $d'$ into a jet, lepton, and missing
energy provides an important signature for this class of models at the LHC.

\section*{Acknowledgments}
This work is partially supported by the ICTP Grant No. AC-80. M. Ashry would like to thank W. Abdallah, A. Elsayed, A. Hammad and A. Moursy  for fruitful discussions.
\section*{Appendix}
To study the boundedness from below, and hence the stability, of the potential (\ref{Scalar Potential}) we use the following theorem \cite{Ping1993109,Kannike:2012pe} to ensure that the matrix of the quartic terms, which are dominant at higher values of the fields, is copositive:
\begin{theorem}[Copositivity Criteria]\label{copcrit}
Let $a\in\mathbb{R},\ b\in\mathbb{R}^{n-1}$ and $C\in \mathbb{R}^{(n-1)\times (n-1)}$. The symmetric matrix $M\in \mathbb{R}^{n\times n}$
\bea M=\left(\begin{array}{cc} a & b^T \\ b & C \end{array}\right)\nn, \eea
is copositive if and only if
\begin{enumerate}
  \item $a\geq0,\ C$ is copositive,
  \item for any nonzero vector $y\in \mathbb{R}^{(n-1)}$, with $y\geq0$, if $b^T y\leq0$, it follows that
$ y^T\left(aC-bb^T\right)y\geq0. $
\end{enumerate}
\end{theorem}
The quartic terms of the potential (\ref{Scalar Potential}) can be written as
\bea\label{quar}
^{4F}V\left(\phi_{1,2}^{0,+},\chi_{L,R}^{0,+}\right)
&=&\lm_1\left(\left|\phi_1^0\right|^4+\left|\phi_1^+\right|^4+\left|\phi_2^0\right|^4+\left|\phi_2^+\right|^4\right)
+\lm_3\left(\left|\chi_L^0\right|^4+\left|\chi_L^+\right|^4+\left|\chi_R^0\right|^4+\left|\chi_R^+\right|^4\right)\nn\\
&+&2\left|\phi_1^0\right|^2\left[\lm_1\left(\left|\phi_1^+\right|^2+\left|\phi_2^+\right|^2\right)+\lm_{12}\left|\phi_2^0\right|^2
+\al_{13}\left(\left|\chi_L^0\right|^2+\left|\chi_R^0\right|^2\right)+\al_{12}\left(\left|\chi_L^+\right|^2+\left|\chi_R^+\right|^2\right)\right]\nn\\
&+&2\left|\phi_1^+\right|^2\left[\lm_1\left|\phi_2^0\right|^2+\lm_{12}\left|\phi_2^+\right|^2+\al_{13}\left(\left|\chi_L^0\right|^2+\left|\chi_R^+\right|^2\right)+\al_{12}\left(\left|\chi_L^+\right|^2+\left|\chi_R^0\right|^2\right)\right]\nn\\
&+&2\left|\phi_2^0\right|^2\left[\lm_1\left|\phi_2^+\right|^2+\al_{12}\left(\left|\chi_L^0\right|^2+\left|\chi_R^0\right|^2\right)+\al_{13}\left(\left|\chi_L^+\right|^2+\left|\chi_R^+\right|^2\right)\right]\nn\\
&+&2\left|\phi_2^+\right|^2\left[\al_{12}\left(\left|\chi_L^0\right|^2+\left|\chi_R^+\right|^2\right)+\al_{13}\left(\left|\chi_L^+\right|^2+\left|\chi_R^0\right|^2\right)\right]\nn\\
&+&2\left|\chi_L^0\right|^2\left(\lm_3\left|\chi_L^+\right|^2+\lm_4\left|\chi_R^0\right|^2+\lm_4\left|\chi_R^+\right|^2\right)
+2\left|\chi_L^+\right|^2\left(\lm_4\left|\chi_R^0\right|^2+\lm_4\left|\chi_R^+\right|^2\right)\nn\\
&+&2\lm_3\left|\chi_R^0\right|^2\left|\chi_R^+\right|^2
-8 \lm_2 Re\left[\phi_1^0 \phi_1^- \phi_2^0 \phi_2^+\right]\nn\\
&+&4(\al_2-\al_3)
Re\left[\left(\phi_1^0 \phi_2^+ + {\phi_2^0}^* \phi_1^+\right) \chi_L^0 \chi_L^- + \left(\phi_2^0 \phi_2^+ + {\phi_1^0}^* \phi_1^+\right) \chi_R^0 \chi_R^- \right],
\eea
where $\al_{12}=\al_1+\al_2,\ \al_{13}=\al_1+\al_3$ and $\lm_{12}=\lm_1+2\lm_2$.
We have
\be
\phi_{1,2}^{0,+}= |\phi_{1,2}^{0,+}|\exp\left[i\theta_{1,2}^{0,+}\right],~~~~~
\chi_{L,R}^{0,+}=|\chi_{L,R}^{0,+}|\exp\left[i\theta_{L,R}^{0,+}\right].
\ee
By the redefinitions of the fields' components,
\be
\phi_1^+\to \phi_1^+ \exp\left[i\left(\theta_1^0-\theta_1^+\right)\right],
\phi_2^0\to \phi_2^0 \exp\left[i\left(\theta_1^0-\theta_2^0\right)\right],
\phi_2^+\to \phi_2^+ \exp\left[-i\left(\theta_1^0+\theta_2^+\right)\right],
\ee
\be
\chi_{L,R}^+\to \chi_{L,R}^+\exp\left[i\left(\theta_{L,R}^0-\theta_{L,R}^+\right)\right],
\ee
we can write
\bea\label{quar1}
^{4F}V\left(\phi_{1,2}^{0,+},\chi_{L,R}^{0,+}\right)&=& X^T~^{4F}V~X-8\lm_2|\phi_1^0||\phi_1^-||\phi_2^0||\phi_2^+|\nn\\
&+&4(\al_2-\al_3)\left[\left(|\phi_1^0||\phi_2^+|+|\phi_2^0||\phi_1^+|\right)|\chi_L^0||\chi_L^+|
+\left(|\phi_2^0||\phi_2^+|+|\phi_1^0||\phi_1^+|\right)|\chi_R^0||\chi_R^+|\right],
\eea
where
\bea
X^T&=&\left(\begin{array}{cccccccc}\left|\phi_1^0\right|^2&\left|\phi_1^+\right|^2&\left|\phi_2^0\right|^2&\left|\phi_2^+\right|^2&\left|\chi_L^0\right|^2&\left|\chi_L^+\right|^2&\left|\chi_R^0\right|^2&\left|\chi_R^+\right|^2\end{array}\right),\\
^{4F}V&=&\left(\begin{array}{cccccccc}
 \lm_1 & ~~\lm_1 & ~~~\lm_{12} & ~~\lm_1 & ~~\al_{13} & ~~~\al_{12} & ~~~\al_{13} & ~~~\al_{12}\\
 \lm_1 & ~~\lm_1 & ~~~ \lm_1 & ~~\lm_{12} & ~~\al_{13} & ~~~\al_{12} & ~~~\al_{12} & ~~~\al_{13}\\
 \lm_{12} & ~~\lm_1 & ~~~\lm_1 & ~~\lm_1 & ~~\al_{12} & ~~~\al_{13} & ~~~\al_{12} & ~~~\al_{13}\\
 \lm_1 & ~~\lm_{12} & ~~~\lm_1 & ~~\lm_1 & ~~\al_{12} & ~~~\al_{13} & ~~~\al_{13} & ~~~\al_{12}\\
 \al_{13} & ~~\al_{13} & ~~~\al_{12} & ~~\al_{12} & ~~~\lm_3 & ~~~\lm_3 & ~~~\lm_4 & ~~~\lm_4\\
 \al_{12} & ~~\al_{12} & ~~~\al_{13} & ~~\al_{13} & ~~~\lm_3 & ~~~\lm_3 & ~~~\lm_4 & ~~~\lm_4\\
 \al_{13} & ~~\al_{12} & ~~~\al_{12} & ~~\al_{13} & ~~~\lm_4 & ~~~\lm_4 & ~~~\lm_3 & ~~~\lm_3\\
 \al_{12} & ~~\al_{13} & ~~~\al_{13} & ~~\al_{12} & ~~~\lm_4 & ~~~\lm_4 & ~~~\lm_3 & ~~~\lm_3
\end{array}\right).
\eea
For the potential (\ref{quar}) to be bounded from below, it must happen that the matrix $^{4F}V$ is copositive and $\lm_2\leq0$ and $\al_2-\al_3\geq0$.
The pseudoscalar Higgs mass (\ref{Amass}) implies that $\mu_3<0$. With the minimization condition (\ref{mca}), both imply that $\lm_4>\lm_3$.
The copositivity implies that the diagonal elements $\lm_1,\lm_3\geq0$. Accordingly, $\lm_4>\lm_3\geq0$.
It is remarkable that the copositivity of the matrix $^{4F}V$ significantly depends on the signs of the parameters $\al_{12},\ \al_{13}$, and $\lm_{12}$.
Here we present the cases depending on these signs:
\begin{enumerate}
\item $\al_{12}\geq0,\ \al_{13}\geq0$, and $\lm_{12}\geq0$:
In this case, the matrix $^{4F}V$ is copositive, and the potential is bounded from below.
\item $\al_{12}\geq0,\ \al_{13}\geq0$, and $\lm_{12}\leq0$:
The copositivity conditions are
\be\label{cop78} \lm_1+\lm_2\geq0,~\lm_1^2+8\lm_1\lm_2+4\lm_2^2\leq0.\ee
We deduce these conditions in detail considering the case assumptions and using Theorem \ref{copcrit}. To make the $8\times8$ matrix $^{4F}V$
be copositive, we shall make that first with the $7\times7$ matrix, $C$, arising from the matrix $^{4F}V$ by eliminating the first row and the first column.
In our case, it is sufficient to stop at this level, since the $6\times6$ matrix, $C_1$, arising from the matrix $^{4F}V$ by eliminating the first two rows and the first two columns is already copositive; being a matrix of nonnegative elements. Now,
\bea
^{4F}V&=&\left(\begin{array}{cc} \lm_1 & b^T\\ b & C \end{array}\right),~~~~~
b^T=\left(\begin{array}{ccccccc}\lm_1 & \lm_{12} & \lm_1 & \al_{13} & \al_{12} & \al_{13} & \al_{12}\end{array}\right),
\\C&=&\left(\begin{array}{cc} \lm_1 & b_1^T\\ b_1 & C_1\end{array}\right),~~~~~
b_1^T=\left(\begin{array}{cccccc}\lm_1 & \lm_{12} & \al_{13} & \al_{12} & \al_{12} & \al_{13}\end{array}\right).
\eea
Let $y_1^T=\left(\begin{array}{cccccc}x_1 & x_2 & x_3 & x_4 & x_5 & x_6\end{array}\right)$ be a vector that satisfies
Theorem \ref{copcrit} requests, {\it i.e.,} a nonzero and a nonnegative vector. Taking $x_2\neq0,\ x_{1,3,...,6}=0$, makes the linear form
$b_1^Ty_1=\lm_{12} x_2\leq0$ and its corresponding quadratic form
\bes
y_1^T\left(\lm_1C_1-b_1b_1^T\right)y_1=-4 \lm_2 (\lm_1+\lm_2) x_2^2.
\ees
Since we have $\lm_2\leq0$, we impose the condition
\be\label{cop7} \lm_1+\lm_2\geq0 \ee
to make the quadratic form $y_1^T\left(\lm_1C_1-b_1b_1^T\right)y\geq0$
and hence as a necessary condition for the copositivity.

Let us assume that $x_i\neq0,\ i=1,...,6$. Then, the linear form
\bes
b_1^Ty_1\leq0 \longleftrightarrow x_2\geq x_2^{\text {min}}=\frac{1}{-\lm_{12}}(\lm_1 x_1+\al_{13} x_3+\al_{12} x_4+\al_{12} x_5+\al_{13} x_6).
\ees
The copositivity condition (\ref{cop7}) makes the corresponding quadratic form be increasing in $x_2$ (for any fixed values of the
other $x_i$'s), and hence we deduce that
\bea
&&y_1^T\left(\lm_1C_1-b_1b_1^T\right)y_1\geq y_1^T\left(\lm_1C_1-b_1b_1^T\right)y_1\Big\vert_{x_2=x_2^{\text {min}}}\nn\\
&&=\frac{\lm_1}{\lm_{12}^2}
\Big[4 \lm_1 \lm_2^2 x_1^2
-2\lm_2\left(\al_{13} \lm_1-\al_{12}\lm_{12}\right)x_1 x_3
-2\lm_2 \left(\al_{12} \lm_1-\al_{13} \lm_{12}\right)x_1 x_4\nn\\
&&+2x_1\left((\al_{12}x_5+\al_{13}x_6)(\lm_1^2+2\lm_1 \lm_2+4\lm_2^2)-2(\al_{13}x_5+\al_{12}x_6)\lm_1 \lm_{12}\right)\nn\\
&&+x_3\left((\al_{13}^2\lm_1-2\al_{12}\al_{13}\lm_{12})(x_3+2x_6)+\lm_{12}^2(\lm_3x_3+2\lm_4x_6)\right)\nn\\
&&+2x_3\left((\al_{12}\al_{13}\lm_1-\lm_{12}(\al_{12}^2+\al_{13}^2))(x_4+x_5)+\lm_{12}^2(\lm_3x_4+\lm_4x_5)\right)\nn\\
&&+x_4\left((\al_{12}^2\lm_1-2\al_{12}\al_{13}\lm_{12})(x_4+2x_5)+\lm_{12}^2(\lm_3x_4+2\lm_4x_5)\right)\nn\\
&&+2x_6\left((\al_{12}\al_{13}\lm_1-\lm_{12}(\al_{12}^2+\al_{13}^2))(x_4+x_5)+\lm_{12}^2(\lm_4x_4+\lm_3x_5)\right)\nn\\
&&+\left(\al_{12}^2\lm_1-2\al_{12}\al_{13}\lm_{12}+\lm_{12}^2\lm_3\right)x_5^2
+\left(\al_{13}^2\lm_1-2\al_{12}\al_{13}\lm_{12}+\lm_{12}^2\lm_3\right)x_6^2\Big].~~~
\label{quad7}~~~~~\eea
By the case assumptions and the copositivity condition (\ref{cop7}), the quadratic form
(\ref{quad7}) is non-negative termwise and the theorem is satisfied.

For the copositivity of the matrix $^{4F}V$, let $y^T=\left(\begin{array}{ccccccc}x_1 & x_2 & x_3 & x_4 & x_5 & x_6 & x_7\end{array}\right)$ be a
nonzero and a non-negative vector. Let $x_{1,2,3}\neq0,\ x_{4,...,7}=0$. Then the linear form
\bes
b^Ty=\lm_1(x_1+x_3)+ \lm_{12} x_2\leq0 \longleftrightarrow x_2\geq x_2^{\text {min}}=\frac{\lm_1}{-\lm_{12}}(x_1+x_3).
\ees
Condition (\ref{cop7}) makes the corresponding quadratic form
\bes
y^T\left(\lm_1C-bb^T\right)y=-4 \lm_2\left(x_1(x_2-x_3)\lm_1+x_2(x_3\lm_1+x_2(\lm_1+\lm_2))\right)
\ees
be increasing in $x_2$ (for any fixed values of $x_{1,3}$), and hence we deduce that
\beas
y^T\left(\lm_1C-bb^T\right)y&\geq&y^T\left(\lm_1C-bb^T\right)y\Big\vert_{x_2=x_2^{\text {min}}}\\
&=&\frac{4\lm_1}{\lm_{12}^2}\left(\lm_1\lm_2^2x_1^2+\lm_2(\lm_1^2+6\lm_1\lm_2+4\lm_2^2)x_1 x_3 +\lm_1\lm_2^2x_3^2\right)\geq0\\
&=&\frac{4\lm_1}{\lm_{12}^2}X_{13}^T M_{13}X_{13},\ \forall x_{1,3},
\eeas
where
\bes
M_{13}=\left(\begin{array}{cc}
  \lm_1\lm_2^2 & \frac{1}{2}\lm_2(\lm_1^2+6\lm_1\lm_2+4\lm_2^2) \\
  \frac{1}{2}\lm_2(\lm_1^2+6\lm_1\lm_2+4\lm_2^2) & \lm_1\lm_2^2
\end{array}\right),~~~~~
X_{13}\left(\begin{array}{c}x_1 \\ x_3\end{array}\right).
\ees
Now, $y^T\left(\lm_1C-bb^T\right)y\Big\vert_{x_2=x_2^{\text {min}}}\geq0$ is equivalent to the copositivity of the matrix
$M_{13}$. Equivalently,
\be\label{cop8}
  \lm_1^2+8\lm_1\lm_2+4\lm_2^2\leq0.
\ee
Now, assume that $x_i\neq0,\ i=1,...,7$. Then, the linear form
\bes
b^Ty\leq0 \longleftrightarrow x_2\geq x_2^{\text {min}}=\frac{1}{-\lm_{12}}(\lm_1 x_1+\lm_1 x_3+\al_{13} x_4+\al_{12} x_5+\al_{13} x_6+\al_{12} x_7).
\ees
As before, conditions (\ref{cop7},~\ref{cop8}) make\beas
y^T\left(\lm_1C-bb^T\right)y&\geq&y^T\left(\lm_1C-bb^T\right)y\Big\vert_{x_2=x_2^{\text {min}}}\geq0,\ \forall x_{1,3,..,6}.
\eeas
Hence, the theorem is satisfied, and, finally, the only imposed conditions for the matrix $^{4F}V$ to be copositive in this case are those in (\ref{cop78}).
The same procedure is followed to extract the copositivity conditions in the following cases:
\item $\al_{12}\geq0,\ \al_{13}\leq0$, and $\lm_{12}\geq0$: The following conditions are necessary for the copositivity:
\bes
\lm_1\lm_3-\al_{13}^2\geq0,~\al_{13}^2(\lm_3-\lm_4)\geq0.
\ees
Since $\lm_4-\lm_3>0$, then we must have $\al_{13}=0$. Finally, in this case, the copositivity conditions are
\be \al_{12}\geq0,\ \al_{13}=0,\ \lm_{12}\geq0. \ee
\item $\al_{12}\leq0,\ \al_{13}\geq0$, and $\lm_{12}\geq0$: The following conditions are necessary for the copositivity:
\bes
\lm_1\lm_3-\al_{12}^2\geq0,~\al_{12}^2(\al_{12}-\al_{13})^2\lm_1^2\left(\lm_3^2-\lm_4^2\right)\geq0.
\ees
Again, either $\al_{12}=0,\ \al_{12}=\al_{13}$, or $\lm_1=0$. But the minimal copositivity conditions in this case are
 \be \al_{12}=0,\ \al_{13}\geq0,\ \lm_{12}\geq0. \ee
\item $\al_{12}\geq0,\ \al_{13}\leq0$, and $\lm_{12}\leq0$: The copositivity conditions are
\be \al_{12}\geq0,\ \al_{13}=0,~\lm_{12}\leq0,\ \lm_1+\lm_2\geq0,~\lm_1^2+8\lm_1\lm_2+4\lm_2^2\leq0.\ee
\item $\al_{12}\leq0,\ \al_{13}\geq0$, and $\lm_{12}\leq0$: The copositivity conditions are
\be \al_{12}=0,\ \al_{13}\geq0,~\lm_{12}\leq0,\ \lm_1+\lm_2\geq0,~\lm_1^2+8\lm_1\lm_2+4\lm_2^2\leq0.\ee
\item $\al_{12}\leq0,\ \al_{13}\leq0$, and $\lm_{12}\geq0$: The following conditions are necessary for the copositivity:
\be \lm_1\lm_3-\al_{12}^2\geq0,~\lm_1\lm_3-\al_{13}^2\geq0,\ \al_{12}^2(\lm_3-\lm_4)\geq0,\ \al_{13}^2(\lm_3-\lm_4)\geq0.\ee
Hence, in this case, the copositivity conditions are
\be \al_{12}=\al_{13}=0,\ \lm_{12}\geq0. \ee
\item $\al_{12}\leq0,\ \al_{13}\leq0$, and $\lm_{12}\leq0$: The copositivity conditions are
\be \al_{12}=\al_{13}=0,\ \lm_{12}\leq0,\ \lm_1+\lm_2\geq0,\ \lm_1^2+8\lm_1\lm_2+4\lm_2^2\leq0. \ee
\end{enumerate}
\bibliographystyle{jhepsty}\bibliography{Bib}
\end{document}